\documentclass[opre,nonblindrev]{informs3}

\OneAndAHalfSpacedXI
\usepackage{bbm}
\usepackage[noend]{algorithmic}
\usepackage{pdflscape}
\usepackage{multirow}
\newtheorem{myprop}{Proposition}
\newtheorem{myproblem}{Problem}
\newcommand{\indicator}[1]{\mathbbm{1}_{\left\{ {#1} \right\} }}
\usepackage{natbib}
 \bibpunct[, ]{(}{)}{,}{a}{}{,}%
 \def\bibfont{\small}%
 \def\bibsep{\smallskipamount}%
 \def\bibhang{24pt}%
\usepackage{hyperref}

\TheoremsNumberedThrough     % Preferred (Theorem 1, Lemma 1, Theorem 2)
\ECRepeatTheorems

\EquationsNumberedThrough    % Default: (1), (2), ...
\MANUSCRIPTNO{}

\begin{document}
%%%%%%%%%%%%%%%%

%\setcounter{page}{1}
\YEAR{2012}% e.g., 2005
%\DOI{10.1287/xxxx.0000.0000}%

\RUNAUTHOR{R Cont and A  Kukanov}

\RUNTITLE{Optimal order placement and routing in limit order markets}

\TITLE{Optimal order placement in limit order markets}

\ARTICLEAUTHORS{%
\AUTHOR{Rama Cont and Arseniy Kukanov}
\AFF{Imperial College London
\&
AQR Capital Management} %, \URL{}}
} % end of the block

\ABSTRACT{%
To execute a trade, participants in electronic equity markets may choose to submit limit orders or market orders across   various  exchanges where a stock is traded. This decision is influenced by characteristics of the order flows  and queue sizes in each limit order book, as well as the structure of transaction fees and rebates across exchanges. We propose a quantitative framework for studying this  {\it order placement} problem  by formulating it as a convex optimization problem. This formulation allows to study how  the optimal order placement decision depends on  the interplay between the state of order books, the fee structure, order flow properties and the aversion to execution risk. In the case of a single exchange, we derive an explicit solution for the optimal split between limit and market orders. For the general case of order placement across multiple exchanges, we propose a stochastic algorithm that computes the optimal routing policy  and study the sensitivity of the solution to various parameters. Our solution  exploits  data on recent order fills across exchanges in the numerical implementation of the algorithm.
}

\KEYWORDS{limit order markets, optimal order execution,  execution risk, order routing, fragmented markets, transaction costs, financial engineering, stochastic approximation, Robbins-Monro algorithm
\vskip 12pt
\noindent {\it First version:} October  2012. {\it Revised:} October 2014.
}

%\noindent {\it Subject classifications}: decision analysis, financial institutions: brokerage/trading, programming: stochastic.
%\vskip 12pt\noindent {\it Area of review}: Stochastic Models

%\KEYWORDS{butter, margarine, silliness} \HISTORY{This paper was
%first submitted on April 12, 1922 and has been with the authors for
%83 years for 65 revisions.}

\maketitle
%%%%%%%%%%%%%%%%%%%%%%%%%%%%%%%%%%%%%%%%%%%%%%%%%%%%%%%%%%%%%%%%%%%%%%

%tableofcontents

\newpage

\section{Introduction}

The trading process in today's automated financial markets can be divided into several stages, each taking place on a different time horizon: portfolio allocation decisions are  made over a time scales of weeks or days  and translate into trades that are executed over time intervals of several minutes to several days through streams of {\it orders} placed at high frequency, sometimes thousands in a single minute \citep{Cont2011}. Existing studies on optimal trade execution have investigated how the execution cost of a large trade may be reduced by splitting it into multiple orders spread in time.
Once this {\it order scheduling } decision is made, one still needs to specify how each individual order should be  placed: this {\it order placement} decision involves the choice of an {\it order type} (limit order or market order), order size and  {\it destination}, when multiple trading venues are available. For example, in the U.S. equity market there are more than ten active exchanges where a trader can buy or sell the same securities. Order placement in a fragmented market is a non-trivial task and brokers offer their clients smart order routing systems in addition to (and often separately from) their suite of trade execution algorithms. We focus here on this {\it order placement} problem: given an order which has been scheduled, choosing an order type --market or limit order-- and which trading venue(s) to submit it to.

Brokers and other active market participants need to make order placement and order routing decisions repeatedly, thousands of times a day, and their outcomes have a significant impact on each participant's transaction cost. An empirical study of proprietary order data from a large execution broker by \citet{Battalio2013} demonstrates that brokers use both limit and marketable orders to execute trades, and the distribution of their orders across trading venues points to strategic routing behavior. Order execution quality is materially affected by order routing choices which recently motivated a number of inquiries by regulators into brokers' ability to optimally place orders on behalf of their clients \citep{USSenate2014,Reuters2014,Bloomberg2014}. The choice between limit and market orders and their routing is important for most market participants, not just brokers. For instance, high-frequency traders can opportunistically provide liquidity with limit orders or demand it with marketable orders and a large group of ``mixed" high-frequency strategies indeed relies on both order types in various proportions (see \citet{Baron2012, Brogaard2014}). At the same time market-makers that simultaneously provide liquidity on multiple trading venues (see \citet{Menkveld2011}) need to take into account the fee and rebate structure and the current state of limit order books at these venues. In aggregate, strategic order placement and routing choices made by a variety of traders shape order flow dynamics in a fragmented market. \citet{Boehmer2006} find evidence that marketable orders are sent to trading venues that provide lower execution costs and \citet{Foucault2008} show that trading fees and the number of available venues affect consolidated market depth. In \citet{Moallemi2011} market orders gravitate towards exchanges with larger posted quote sizes and low fees, while limit orders are submitted to exchanges with high rebates and lower execution waiting times. The importance of order placement and routing decisions on trading performance of individual market participants justifies a more detailed modeling of  order placement and routing decisions.

\subsection{Literature review}

Theoretical studies of limit order markets have previously considered the choice of market/limit order type and execution venue using  stylized models of trader behavior. In \citet{Foucault1999, foucault05, rosu09} traders can submit one limit or market order for one trading unit to a single exchange. There are no order queues and bid-ask spreads are determined in a competitive equilibrium. The choice between a limit and a market order by each trader depends on his exogenous patience parameter and the spread that he observes upon arrival. In \citet{parlour98} traders choose to place a single-unit limit or market order based on their patience and a probability of limit order execution. There is a single exchange and traders form queues at exogenously given bid/ask prices. \citet{Foucault2008} present a model where limit order traders form queues at two exchanges based on a profit break-even condition and then a broker can choose to send market orders to one or both venues. These models describe market dynamics and trader behavior in equilibrium, but require strong simplifying assumptions regarding each trader's order placement and routing behavior. Our analysis complements existing literature by studying the order placement decision itself and providing insights into its structure. For instance, most theoretical models assume that market participants need to execute one trading lot, whereas in practice traders often need to place larger orders (e.g. the average limit order size in the U.S. equity market is close to 400 shares). We find that the size of an order to be placed is actually one of the most interesting input variables. It largely determines the optimal mix of market and limit orders as well as traders' sensitivity to exchange fees and rebates. Stylized models of trader behavior that previously appeared in theoretical literature translate into simple, binary choices of order type and venue based on one or two variables (e.g. bid-ask spread and trader patience). Our analysis shows that order placement and routing decisions are more complex and need to be based on a variety of factors. Comparing the performance of our optimal order placement strategy to simpler rules-of-thumb for a range of parameters we find that simple heuristics are too inflexible and often result in large transaction costs.
This further motivates a need for a detailed analysis of order placement and routing decisions, but there are few studies dedicated to this topic.

A reduced-form model for routing an infinitesimal limit order to one destination is presented in \cite{Moallemi2011}, \cite{almgrenharts} propose a market order routing algorithm in presence of hidden liquidity, while \cite{Ganchev2010} and \cite{Laruelle2009} propose numerical algorithms to optimize order executions across multiple dark pools, where supply/demand is unobserved.  To the best of our knowledge this work is the first to provide a detailed treatment of trader's order placement decision in a multi-exchange market.

Our results also complement the literature on optimal trade execution. Early work on this subject \citep{bertsimas98,almgren00} focused on the scheduling of orders in time but did not explicitly model the process whereby each order is filled. More recent formulations have tried to incorporate some elements in this direction. In one stream of literature \citep{obizhaeva05,Alfonsi2010,Predoiu2011} traders are restricted to using only market orders whose execution costs are given by an idealized order book shape function. Another approach has been to model the process through which an order is filled as a dynamic random process \citep{Cont2011, Cont2011b} leading to a formulation of the optimal execution problem as a stochastic control problem. This formulation has been studied in various settings with limit orders \citep{Bayraktar2011, Gueant2012} or limit and market orders \citep{Guilbaud2012, Huitema2012, Guo2013, Li2013} but its complexity  makes it computationally intractable unless restrictive assumptions are made on price and order book dynamics. For example, these studies commonly assume that a trader places a single limit order for one unit at a time and its execution probability is given by a simplified function of distance to best quotes. In our formulation limit order fills are based on order quantity, queue position and order flows generated by other traders, as they are in actual limit order books. Existing approaches to optimal trade execution do not consider the option of simultaneously placing limit orders on multiple exchanges, the queue position of individual orders and the possibility of receiving partial fills, all of which play a central role in our model.

\subsection{Summary of contributions}
In the present work, we adopt a  more tractable approach which is closer to order routing  methods used in  practice, by separating the {\it order placement} decision from the scheduling decision: assuming that the trade execution schedule has been specified, we focus on the task of filling the scheduled batch of orders (a trade ``slice") by optimally distributing it across trading venues and order types. Decoupling the scheduling problem from the order placement problem is closer to market practice (see e.g. \cite{almgrenharts}) and leads to a more tractable approach allowing us to incorporate some realistic features which matter  for order placement decisions, while conserving analytical tractability. Although simultaneous optimization of order timing, type and routing decisions is an interesting problem, it also appears to be intractable and its solution would likely omit details that are important for either order placement or scheduling. 

Our key contribution is a quantitative formulation of the order placement problem which illustrates how various factors - the size of an order to be placed, lengths of order queues across exchanges, statistical properties of order flows in these exchanges, trader's execution preferences, and the structure of liquidity rebates across trading venues -  blend into an optimal allocation of limit orders and market orders across available trading venues. When only one exchange is available for execution, this order placement problem reduces to the problem of choosing an optimal split between market orders and limit orders. We derive an explicit solution for this problem and analyze its sensitivity to the order size, the trader's urgency for filling the order and other factors. In a case of multiple exchanges we also derive a characterization of the optimal order allocation across trading venues. Finally, we propose a fast and flexible numerical method  for solving the order placement problem in a general case and demonstrate its efficiency through examples.

An important aspect of our framework is to account for   {\it execution  risk}, i.e. the risk of not filling an order. Previous studies focus on the risk of price variations over the course of a trade execution \cite{almgren00,huberman2005} but assume that orders are always filled. However execution risk is a major concern for strategies that involve limit orders \citep{Harris1996}. When it is costly to catch up on the unfilled portion of the order, we find that the optimal allocation shifts from limit to market orders. Although market orders are executed at a less favorable price,  it becomes optimal to use them  when the  execution risk is a primary concern - for example when execution is subject to a deadline or when traders have time-sensitive information about returns.

Optimal limit order sizes are strongly influenced by queue position that they can achieve at each exchange and by distributions of order outflows from these queues. For example, if the queue size at one of the exchanges is much smaller than the expected future order outflow there, it is optimal to place a larger limit order on that exchange.  \cite{Moallemi2011} argue that such favorable limit order placement opportunities vanish in equilibrium due to competition and strategic order routing of individual traders. However their empirical results also show that short-term deviations from the equilibrium are a norm, and can therefore be exploited in our optimization framework.

Our   order placement model brings new insights into the structure of order placement decisions. We find that the targeted execution size plays an important role due to a {\it bounded execution capacity} of limit orders. Relatively small quantities can be executed with a high probability using limit orders placed just at the cheapest exchange. Faced with progressively larger quantities a trader realizes that filling the entire amount with a single order at the cheapest exchange is unlikely and is forced to place orders on more expensive venues. Interestingly, for relatively large quantities the optimal order allocation becomes practically insensitive to rebates as the non-execution risk outweighs the cost of placing limit orders on more expensive exchanges. To fill even larger quantities a trader needs to start using market orders. After some point the optimal market order size increases linearly with the targeted execution quantity while limit order sizes remain bounded at a certain value. As a result of this feature, a larger fraction of the targeted quantity is executed with market orders when that quantity is relatively large.

Another important aspect that appears in our analysis is the tendency of an optimal order allocation to place more orders than it needs to fill. This {\it overbooking} behavior is due to the possibility of receiving partial fills and the availability of multiple exchanges in our model. Instead of placing one big limit order, a trader can place more orders to all available venues, collect their fills and cancel the excess orders afterwards. This results in a reduction of non-execution risk because limit order fills are not perfectly correlated. Since each venue adds new limit order execution opportunities, overbooking becomes more prominent as the number of available venues increases. This  may explain the so-called ``phantom liquidity" observed in the U.S. equity market, i.e. a large amount of limit orders that are quickly canceled before market order traders can execute against them. Assuming that market-makers and other limit order traders indeed post multiple orders across exchanges and cancel all substitute orders once one of them is filled, these ``phantom" orders may represent rational attempts to reduce risk rather than a manipulative practice. Further extrapolating this overbooking behavior to all limit order traders, our model would predict an increase in consolidated depth with each additional trading venue. This intuition is similar to \citet{Foucault2008}, although in their model limit orders are placed only by profit-maximizing market-makers that would not post orders at an expensive venue. In our model, overbooking (and thus an increase in consolidated depth) can occur despite  high exchange fees, which suggests that consolidated depth  increases with the introduction of additional trading venues, as long as these venues allow limit order traders to diversify their execution risk.

\subsection{Outline}

Section \ref{sec.routing} describes our formulation of the order placement problem and presents conditions for the existence of an optimal order placement. In Section \ref{sec.1exchg} we derive an optimal split between market and limit orders for a single exchange. Section \ref{sec.2exchg}  analyzes the general case of order placement on multiple trading venues. Section \ref{sec.sa} presents a numerical algorithm for solving the order placement problem in a general case and studies its convergencee properties. Section \ref{ssec.compstat} analyses the structure of order placement decisions through comparative statics. Section \ref{ssec.tick} presents an application of our method to historical tick data and Section \ref{sec.conc} concludes.

\newpage
\section{The order placement problem}\label{sec.routing}
\subsection{Decision variables}

Consider a trader who needs to buy $S$ shares of a stock within a short time interval $[0,T]$. The deadline $T$ may be a fixed time horizon such as 1 minute, or a random stopping time, for instance triggered by price changes or cumulative trading volume. The execution target $S$ is assumed to be relatively small - it is a slice of the daily trade that the trader expects to fill during $[0,T]$. Nevertheless, this slice $S$ is usually larger than a single trading lot because limit orders may have to queue for execution for several seconds or even minutes. Our objective is to define a meaningful framework in which the trader can compare alternative approaches to executing this trade slice $S$, for example choose between sending a limit order to a single exchange, splitting it in some proportion across $K$ exchanges or using a combination of market and limit orders. We assume the following two-step execution strategy - at time 0 the trader may submit $K$ limit orders for $L_k\geq 0$ shares to exchanges $k=1,\dots,K$ and also market orders for $M\geq 0$ shares. 

At time $T$ if the total executed quantity is less than $S$ the trader also submits a market order to execute the remaining amount. The trader's {\it order placement} decision  is thus  summarized by a vector $X~=~(M,L_1,\dots,L_K)\in\mathbb{R}^{K+1}_+$ of order sizes. The components of $X$ are non-negative (only buy orders are allowed) and we assume that the trader has no other (e.g. pre-existing) orders in the market.

\subsection{Order execution}

We will assume for simplicity that a market order of any size up to $S$ can be filled immediately at any single exchange. Thus a trader chooses the cheapest venue for his market orders.

Limit orders with quantities $(L_1,\dots,L_K)$ join queues of $(Q_1,\dots,Q_K)$ orders in $K$ limit order books, where $Q_k\geq 0$. As a simplification we assume that all of these limit orders have the same price - the highest bid price across venues, i.e. the National Best Bid. The case $Q_k=0$ is allowed in our model and corresponds to placing limit orders inside the bid-ask spread at one of the venues.

\begin{figure}[ht]
\noindent
\begin{center}
{
\includegraphics[width=110mm]{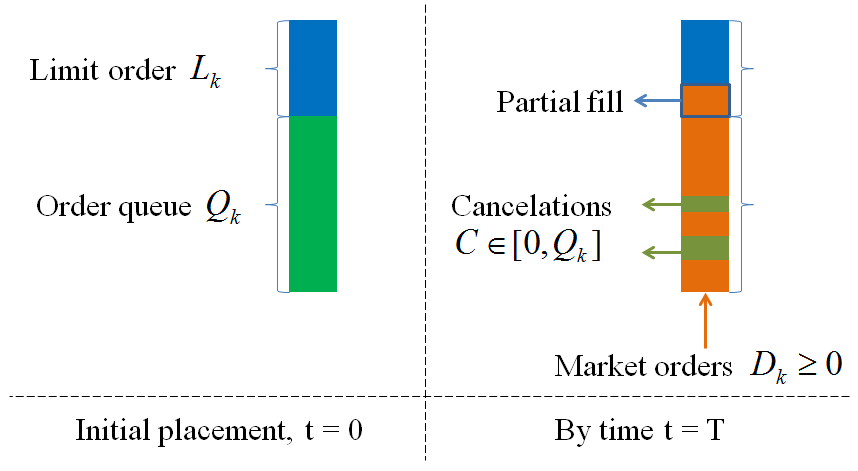}
}
\end{center}
\caption{Limit order execution on exchange $k$ depends on the order size $L_k$, the queue $Q_k$ in front of it, total sizes of order cancelations $C_k$ and marketable orders $D_k$, specifically on $\xi_k=C_k+D_k$.}
\label{fig.queue_fill}
\end{figure}

Denote by $(x)_+=\max(x,0)$. Using the assumption that limit orders are not modified before time $T$, we can explicitly calculate their filled amounts (full or partial) as a function of their initial queue position and future order flow:

$$\min(\max(\xi_k-Q_k,0),L_k)=(\xi_k-Q_k)_+-(\xi_k-Q_k-L_k)_+,\quad k=1,\dots,K$$

\noindent where $\xi_k$ is a total outflow from the front of the $k$-th  queue. The order outflow $\xi_k$ consists of order cancelations that occurred before time $T$ from queue positions in front of an order $L_k$, and marketable orders that reach the $k$-th exchange before $T$. The mechanics of limit order fills in a FIFO queue are illustrated in Figure \ref{fig.queue_fill}.

We note that limit order fill amounts are random because they depend on queue outflows $\xi=(\xi_1,\dots,\xi_K)$ during $[0,T]$, modeled as a random variable with a distribution $F$. Our formulation  also does not require specifying the distribution of $\xi$.
The {\it total amount} of shares $A(X,\xi)$ bought by the trader during  $[0,T)$ can be written as a function of the initial order allocation $X$ and intermediate queue outflows $\xi$:
\begin{equation}\label{optim.amt}
A(X,\xi)=M+\sum_{k=1}^K\left((\xi_k-Q_k)_+-(\xi_k-Q_k-L_k)_+\right)
\end{equation}
If the executed amount $A(X,\xi) < S$ we assume that the trader submits a market order at time $T$ ensuring the execution of this trade slice $S$.

\subsection{Cost function}

The trader's objective is to minimize the sum of explicit and implicit costs associated with order execution. There is a variety of explicit costs that vary by exchange. Most U.S. equity exchanges charge fees to market order traders for consuming liquidity and pay rebates to limit order traders for providing it. In contrast, some {\it inverse} exchanges pay traders for marketable orders and charge for providing liquidity. For example, at the time of writing the U.S. equity exchange Direct Edge EDGA had a negative rebate $-\$0.0006$ per share for passive orders and a  negative fee $-\$0.0004$ per share for marketable orders\footnote{See \citet{Battalio2013} for a   comprehensive summary of fees charged by U.S. equity exchanges.}.  These fees and rebates are economically significant and similar in magnitude to bid-ask spread costs ($\$0.0050$ per share for liquid U.S. stocks). Exchanges outside of U.S. have adopted similar fee/rebate pricing structures (e.g. BATS Chi-X Europe), or special rebate programs for liquidity providers (e.g. Singapore Stock Exchange, BMF-BOVESPA).

Implicit execution costs include adverse selection and market impact. Adverse selection is reflected in the correlation between limit order executions and price changes (\cite{glosten85}). For example, after a buy limit order is filled prices tend to go down below its limit price creating an immediate loss for a limit order trader. The magnitude of adverse selection losses varies by exchange, and venues with high rebates are typically exposed to more adverse selection (see \cite{Battalio2013}) which can be explained in a rational equilibrium model of \cite{Moallemi2011}. Small but consistent losses on limit order fills can accumulate to a significant adverse selection cost over time, which motivates us to use effective rebates $r_k = r^{e}_k + AS_k$, where $r^{e}_k$ are rebates set by exchanges and $AS_k$ are exchange-specific penalties for adverse selection. In practice these penalties are often chosen empirically as average returns measured over a short interval following a limit order execution.

Using the mid-quote price as a benchmark, we calculate execution costs relative to mid-quote for an order allocation $X=(M,L_1,\dots,L_K)$ as:

\begin{equation}\label{optim.cost}
(h+f)M-\sum_{k=1}^K(h+r_k)((\xi_k-Q_k)_+-(\xi_k-Q_k-L_k)_+),
\end{equation}
\noindent where $h$ is one-half of the bid-ask spread at time 0, $f$ is a fee for market orders and $r_k$ are effective rebates for limit orders on exchanges $k=1,\dots,K$.

In addition to average adverse selection losses $AS_k$ incurred on filled limit orders, a trader may experience a shortfall due to unfilled limit orders. In the event $A(X,\xi)<S$ the trader has to purchase the remaining $S-A(X,\xi)$ shares at time $T$ with a costly market order. Adverse selection implies that conditionally on this event prices have likely increased and the cost of market orders at time $T$ is higher than their cost at time 0, i.e. $\lambda_u>h+f$. Alternatively, in the event $A(X,\xi)>S$ the prices likely decreased even more than after an average limit order fill, and $\lambda_o$ measures this additional adverse selection cost. To capture this {\it execution risk} we include, in the objective function,  a penalty  for violations of  target quantity in both directions:
\begin{equation}\label{optim.tarpen}
\lambda_u\left(S - A(X,\xi)\right)_+
+\lambda_o\left(A(X,\xi)-S\right)_+ ,
\end{equation}
\noindent where $\lambda_u\geq 0,\lambda_o\geq 0$ are marginal penalties in dollars per share for, respectively falling behind or exceeding the execution target $S$. In addition to adverse selection, the penalties $\lambda_u,\lambda_o$ may reflect trader's private execution preferences. Generally, a trader can tolerate some differences between $A(X,\xi)$ and $S$ because $S,~T$ are fractions of the overall trade quantity and time horizon. The penalties need not be symmetric - a trader with a positive forecast of short-term returns within the period $T$ has a larger opportunity cost and may set $\lambda_u>\lambda_o$.

We also include market impact as a function of the volume of submitted orders. The target quantity $S$ is assumed to be small so orders $(M,L_1,\dots,L_K)$ may have little immediate impact on prices in the interval $[0,T]$. However, this impact may accumulate over the course of trading. Accounting for average impact costs is important: it penalizes order placement strategies that submit too many orders or orders that are too large. Empirical studies show that both market and limit orders affect prices, and the average impact of small orders can be well approximated by a linear function (\cite{ContKukanov2013}),  as in \cite{Kyle1985}. We assume that the impact cost is paid on all orders placed at times $0$ and $T$, irrespective of whether they are filled, leading to the following total impact:
\begin{equation}\label{optim.impact}
\theta\left(M+\sum_{k=1}^KL_k+\left(S - A(X,\xi)\right)_+\right),
\end{equation}
\noindent where $\theta>0$ is the impact coefficient.

Adding these different terms we obtain:
\begin{definition}[Cost function]
The cost function  is defined as the sum of explicit and implicit execution costs:
\begin{align}
v(X,\xi):&=(h+f)M-\sum_{k=1}^K(h+r_k)\left((\xi_k-Q_k)_+-(\xi_k-Q_k-L_k)_+\right) \notag\\
&+\theta\left(M+\sum_{k=1}^KL_k + \left(S - A(X,\xi)\right)_+\right) + \lambda_u\left(S - A(X,\xi))\right)_+
+\lambda_o\left(A(X,\xi)-S\right)_+ \label{eq.V}
\end{align}
\end{definition}
It involves the following ingredients:
\begin{itemize}
\item Execution objectives: target quantity $S$, time horizon $T$
\item Trading costs: half of bid-ask spread $h$, market order fee $f$ and effective limit order rebates $r_k$, market impact coefficient $\theta$, penalties for under- and overfilling the target $\lambda_u, \lambda_o$
\item Market configuration: number of exchanges $K$, limit order queues $Q_k$.
\end{itemize}
\subsection{Optimal order placement problem}

We can now formulate the search for an optimal order placement as a cost minimization problem:
\begin{myproblem}[Optimal order placement problem]\label{prob.optimalplacement}
An {\it optimal order placement} is a vector $X^*\in\mathbb{R}^{K+1}_+$ solution of
\begin{equation}\label{optim.set}
\min_{X\in\mathbb{R}^{K+1}_+}V(X)
\end{equation}
\noindent where
\begin{align}
V(X)= \mathbb{E}[ v(X,\xi) ]= \int_{\mathbb{R}^d} F(dy) v(X,y)
\end{align}
\noindent is the expected execution cost for the allocation $X$ and the expectation is taken with respect to the distribution  $F$ of  order outflows $(\xi_1,...,\xi_K)$ at horizon $T$.
\end{myproblem}
\noindent 
The output is an order allocation $X^\star=(M^\star,L^\star_1,\dots,L^\star_K)$  consisting of a market order quantity $M^\star$ and limit order quantities $L^\star_1,\dots,L^\star_K$ which minimizes the expected execution cost over $[0,T]$.

\subsection{Discussion}

Before proceeding to results we discuss some of the important assumptions in our model and their implications.

{\it Continuous decision variables}: In reality orders are integer multiples of a share; however batch sizes are often large and one can neglect in first instance the granularity of orders and optimize over $X\in \mathbb{R}^{K+1}_+$, then round to number of shares in the last step. This procedure, which corresponds to the convex relaxation of the underlying integer optimization problem \citep{williamson2011}, is indeed the standard approach used in the optimal execution literature \cite{almgren00,Alfonsi2010,Bayraktar2011,Guilbaud2012}.

{\it Static optimization problem}: In a dynamic setting, one would need to solve Problem \ref{prob.optimalplacement} one-step ahead, using the conditional distribution of $\xi$ if known:
\begin{equation} 
V(t_k,X_k^*)=\min_{X\in\mathbb{R}^{K+1}_+}  \mathbb{E}[ v(X,\xi_t)|{\cal F}_t ]
\end{equation}
so insights from Problem \ref{prob.optimalplacement} are useful for understanding the dynamic version of the problem. Problem \ref{prob.optimalplacement} may be seen as the stationary/ ergodic version of the problem, in which one considers the average cost over many trades, rather than the one-step-ahead execution cost for a single order placement. An alternative approach to order placement based on a constrained optimization problem is discussed in the Appendix.

{\it Execution certainty for market orders}: This assumption appears to be valid as long as $S$ is of the same magnitude as the prevailing market depth (roughly 600 shares for an average US stock). Our results are easily extended, at the expense of additional notation, to a case where $S$ is large but still can be filled with multiple market orders with progressively higher prices or fees. For example, consider a case when there are only $D_1<S$ shares available at the cheapest exchange with a fee $f_1$, but additional shares are available at a more expensive venue with a fee $f_2$. The trader can fill $S$ shares by sending two market orders and their total explicit cost becomes a piece-wise linear function of total size: $f_1\min(S,D_1) + f_2\max(S-D_1,0)$. If $S$ is even larger, one may add more terms to this function - additional exchanges or deeper levels in the order book - by suitably increasing their marginal costs. This generalization remains tractable as long as the cost function remains convex (e.g. piece-wise linear). However, we note that market order routing is itself a non-trivial problem. Practical solutions need to take into account hidden liquidity, market data speed and trading venue geography which affects latency. These considerations are outside of the scope for our paper that focuses primarily on limit orders and their execution in multiple order queues.

{\it Limit order placement}: We assume that limit orders $L_1,\dots,L_K$ are all placed at the same price - the best prevailing quote. Effectively the pricing decision is narrowed to two options - a limit order at the best quote or a marketable order. This allows us to study limit order execution in more detail focusing on a queue of orders within a specific price level. The assumed choice  appears to be the most interesting case for applications, since in practice brokers submit the majority of limit orders at the best bid and ask prices - see \cite{Cao2008, Battalio2013} for recent statistics.

{\it Exogenous spread}: The spread $h$ is exogenously set in our model, which is related our limit order placement assumption - the trader joins an existing best quote but does not improve it. Although this assumption appears to be restrictive from a theoretical viewpoint, in practice many liquid assets consistently trade with a bid-ask spread equal to a single price increment, which is exogenously set by exchanges or regulators. When the spread narrows to a single price increment, traders cannot improve the existing quote and need to queue with their limit orders, as described in our model.

{\it Market impact coefficient}:  The coefficient $\theta$ could theoretically be different for market and limit orders, as well as for orders sent to different exchanges. However, empirical studies show that market impact differences between limit and market orders are small (\cite{eisler10,Mastromatteo2013}). We also note that market impact occurs over time horizons that are much longer than those involved in order placement (days as opposed to minutes or seconds), suggesting that $\theta$ - the marginal impact of a single share - is much smaller in magnitude than bid-ask spread $h$ and explicit costs $f,r_k$. For example, consider a trade to buy 5\% of daily volume in a liquid stock. Recent empirical studies \citep{almgren05, Mastromatteo2013} find that the impact of such trade is in a range of 1-6\% of daily volatility, corresponding to 2-12 basis points for a stock with 40\% annualized volatility. In other words, the average change in a stock price due to this trade is 2-12 basis points over the course of multiple hours when the trade is executed. Costs associated with order placement decisions have similar magnitude but are incurred over much smaller time horizons. For example, the difference between using a market order and a limit order for a U.S. stock priced at 30 dollars per share and a spread of 1 penny can be more than 5 basis points after accounting for liquidity fees and rebates.

\subsection{Existence of solutions}

We begin by stating certain economically reasonable restrictions on parameter values, that will be assumed for all of our results.

\noindent {\bf A1}: $\underset{k}{\min}\{r_k\}+h > 0$. Interpretation: even if some effective rebates $r_k$ are negative, limit order executions let the trader earn a fraction of the bid-ask spread.

\noindent {\bf A2}: $\lambda_o > h+\underset{k}{\max}\{r_k\}$ and $\lambda_o > -(h+f)$. Interpretation: the trader has no incentive to exceed the target quantity $S$.

\noindent {\bf A3}: $\lambda_u > h+f$. Interpretation: market orders sent at time 0 are less expensive compared to market orders that are sent at time $T$ conditionally on not being able to fill the target before $T$.

%\vskip 6pt

\noindent Although negative rebate values are possible, in the U.S. they are smaller than the smallest possible value of $h=\$0.005$, justifying our assumptions A1 and A2. Assumption A3 is motivated by adverse selection and explained above.

Our first result, whose proof is given in  the Appendix, shows that it is not optimal to submit limit or market orders that are {\it a priori} too large or too small (larger than the target size $S$ or whose sum is less than $S$):
\begin{myprop}\label{prop.lubound} Under assumptions A1-A3, any optimal order allocation belongs to the set:
 $$\mathcal{C}=\left\{X\in\mathbb{R}^{K+1}_+\Bigm| 0\leq M\leq S,~~0\leq L_k\leq S-M, k=1,\dots,K, ~~M+\sum_{k=1}^K{L_k}\geq S\right\}.$$
\end{myprop}
Proposition \ref{prop.lubound} shows that it is never optimal to overflow the target size $S$ with a single order, but it may be optimal to exceed the target $S$ with the sum of order sizes $M+\sum_{k=1}^K{L_k}$. The penalty function (\ref{optim.tarpen}) effectively implements a soft constraint for order sizes and focuses the search for an optimal order allocation to the set $\mathcal{C}$. Specific economic or operational considerations could also motivate adding hard constraints to problem \eqref{optim.set}, e.g. $M=0$ or $\sum_{k=1}^K{L_k}=S$. Such constraints can be easily included in our framework but absent the aforementioned considerations we do not impose them here.

The existence of an optimal solution is then guaranteed:
\begin{myprop}\label{prop.objfun}
Under assumptions A1-A3, $V:\mathbb{R}^{K+1}_+\mapsto \mathbb{R}$  is convex, bounded from below and has a global minimizer%\footnote{It is actually sufficient to require $s+\min_k\{r_k\}\geq 0,~~\lambda_o > s+\max_{k}\{r_k\},~~s+f\geq 0$}
 $~X^\star\in\mathcal{C}$.
\end{myprop}
We note that the optimal solution may be non-unique, i.e. there could be an optimal ``plateau" depending on the distribution of $\xi$.

%\noindent \redct{Remove this? Think of a unique minimal solution.} . %The problem (\ref{optim.set}) is a convex stochastic optimization problem, and it can be solved numerically by stochastic approximation methods. However, its tractable structure also allows us to establish some analytical results.

\section{Optimal order allocation}
\subsection{Choice of order type: limit orders vs market orders}\label{sec.1exchg}

To highlight the tradeoff between limit and market order executions in our optimization setup, we first consider the case when the stock is traded on a single exchange, and the trader has to choose an optimal split between limit and market orders. This special case is also economically important because many financial assets (e.g. futures contracts, emerging market equities) in fact trade on a single exchange.  
\begin{myprop}[Single exchange: optimal split between limit and market orders]\label{prop.1exchg}
%Assume that $\xi$ has a continuous distribution 
Under assumptions A1-A3:
\begin{enumerate}
\item[(i)] If $\lambda_u\leq\underline{\lambda_u}=\dfrac{2h+f+r}{F(Q+ S)}-(h+r+\theta)$, $(M^{\star},L^{\star})=(0,S)$ is an optimal allocation.
\item[(ii)]
If $\lambda_u\geq\overline{\lambda_u}=\dfrac{2h+f+r}{F(Q)}-(h+r+\theta)$, $(M^{\star},L^{\star})=(S,0)$ is an optimal allocation.
\item[(iii)]
If $\lambda_u\in(\underline{\lambda_u},\overline{\lambda_u})$, an optimal allocation is a mix of limit and market orders, given by
\begin{equation}\label{sol.1exchg}
\left\{
\begin{aligned}
& M^{\star}=S - F^{-1}\left(\frac{2h+f+r}{\lambda_u+h+r+\theta}\right)+Q, \\
& L^{\star}=F^{-1}\left(\frac{2h+f+r}{\lambda_u+h+r+\theta}\right)-Q,
\end{aligned}
\right.
\end{equation}
\noindent where $F(x)=\mathbb{P}(\xi\leq x)$ is the distribution   of the bid queue outflows $\xi$ and $F^{-1}$ its left-inverse.
\end{enumerate}
\end{myprop}

\begin{figure}[h!]
\noindent
\begin{center}
{
\includegraphics[width=120mm]{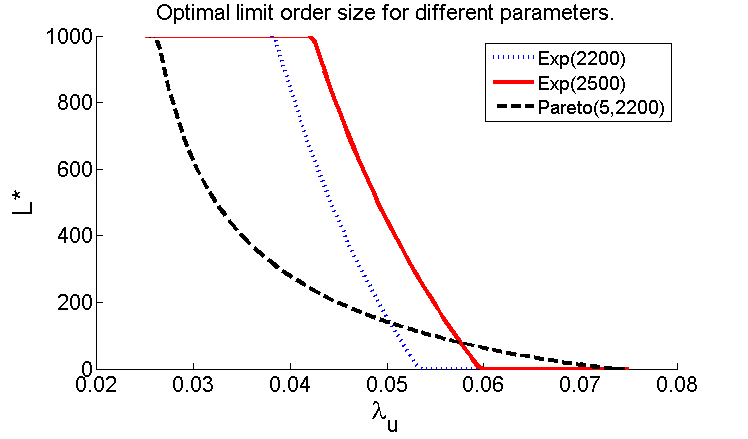}
}
\end{center}
\caption{Optimal limit order size $L^{\star}$ for one exchange. The parameters for this figure are: $Q = 2000,S = 1000, h = 0.02, r = 0.002, f = 0.003, \theta=0.0005$. Colors correspond to different order outflow distributions - exponential with means 2200 and 2500 and Pareto with mean 2200 and a tail index 5.}
\label{fig.Sl_onex}
\end{figure}

\noindent In the case of a single exchange, Proposition \ref{prop.lubound} implies that $M^{\star}+L^{\star}=S$, i.e. the trader does not oversize orders. As a consequence there is no risk of exceeding the target size and $\lambda_o$ does not affect the optimal solution. The trader is only concerned with the risk of falling behind the target quantity, and balances this risk with a fee, rebate and other parameters.  The parameter $\lambda_u$ is an opportunity cost of not filling the target size, and higher values of $\lambda_u$ lead to smaller limit order sizes, as illustrated on Figure \ref{fig.Sl_onex}. Given that $M^{\star}+L^{\star}=S$, the optimal market order size increases with an increase in $\lambda_u$.

The optimal split between market and limit orders depends on the distribution of outflow (execution+cancellation) through its quantile at the level $\frac{2h+f+r}{\lambda_u+h+r+\theta}$: this last formula expresses a tradeoff between marginal costs and savings from a market order. Increasing $M$ by 1 share immediately increases the cost by $h+f$. Since $M+L=S$, the trader also needs to reduce his limit order size. This will increase the cost as the trader loses $h+r$ in potential savings, if that limit order is assumed to fill. However if that limit order does not fill, the trader will need to catch up paying additional $\lambda_u + \theta$ and not realizing any of the savings. The numerator $2h+f+r$ reflects costs that the trader accepts by increasing $M$ by 1 share, assuming his limit order will fill. The denominator reflects market order benefits - if a limit order does not get a fill, the trader does not need to pay an additional $\lambda_u + \theta$ and can forfeit $h+r$ in unrealized savings.

The optimal limit order size decreases with $\lambda_u$ as it becomes more expensive not to fulfill the target, and increases with $f$ as market orders become more expensive.

Another interesting feature is that within the range $\lambda_u\in(\underline{\lambda_u},\overline{\lambda_u})$, $L^{\star}$ is fully determined by $Q$, $F$ and cost parameters, while $M^{\star}$ increases with $S$. The consequences of this solution feature are important. As the target size $S$ increases, a larger fraction $\frac{M^{\star}}{S}$ of that size is executed with a market order. To fill a large limit order a trader would need an improbably large queue outflow $\xi$, so to accommodate larger quantities one needs to resort to market orders. This {\it bounded capacity} feature of limit orders also appears in our solutions for multiple exchanges. For example, as the number of available exchanges $K$ increases, the overall prospects of filling limit orders improve (i.e. their capacity increases) and the share of market orders $\frac{M^{\star}}{S}$ decreases. This is related to the ``crowding-out" effect studied by \citet{parlour98} and empirically confirmed by \citet{Cao2008}. Traders that observe short queues take advantage of that by submitting limit orders, which increases queues and causes subsequent traders to use market orders.

It is important to note that the optimal solution $(M^\star, L^\star)$ depends on the full distribution $F(\cdot)$ of $\xi$ and not just on its mean. Limit orders are filled when $\xi\geq Q+L$, so the tail of $F(\cdot)$ affects order executions and is an important determinant of the optimal order allocation. Figure \ref{fig.Sl_onex} compares order allocations for exponential and Pareto distributions   with equal means. 

%\footnote{The effect of parameters $r, s$  on $M^{\star}$ and $L^{\star}$ is ambiguous when there is only one exchange and it depends on the values of other parameters. Differentiating $L^{\star}$ with respect to $r,s$ we see that $L^{\star}$ increases with $r$ if $\lambda_u>s+f$, otherwise it decreases with $r$. Also, $L^{\star}$ increases with $s$ if $2\lambda_u+r>f$, otherwise it decreases with $s$.}.
%\newpage

\subsection{Optimal routing of limit orders across multiple exchanges}\label{sec.2exchg}

When multiple trading venues are available, dividing a target quantity among them reduces the risk of not filling an order and improves execution quality, and one may even consider sending $S$ shares to each exchange. However, sending too many orders leads to excessive market impact and an undesirable possibility of over-trading. Proposition \ref{prop.foc} gives optimality conditions for an order allocation $X^{\star}=(M^{\star},L^{\star}_1,\dots,L^{\star}_K)$:
\begin{myprop}\label{prop.foc}
Assume A1-A3 hold, the distribution of $\xi$ is continuous, $\underset{k}{\max}\left\{F_k(Q_k+S)\right\}<1$ and $\lambda_u<\underset{k}{\max}\left\{\dfrac{2h+f+r_k}{F_k(Q_k)}-(h+r_k+\theta)\right\}$. Then:
\vskip 6pt
\begin{enumerate}
\item Let $p_0=\mathbb{P}\left(\xi_1\leq Q_1,...,\xi_k\leq Q_K\right)$. If 
$$\qquad \lambda_u\geq\frac{2h+f+\underset{k}{\max}\{r_k\}}{p_0}-(h+\underset{k}{\max}\{r_k\})$$ then optimal order placement strategy involves market orders: $M^\star>0$.
\item If cost savings and fill probability on exchange $j$ overweigh market impact :
$$\qquad (h+r_j-\lambda_o)\mathbb{P}\left(\xi_j>Q_j\right) > \theta$$
then any optimal order placement   involves limit orders on exchange $j$: $L_j^\star>0$.
\item If previous   assumptions hold for all exchanges $j=1,\dots,K$,  $X^\star\in\mathcal{C}$ is an optimal allocation if and only if
\begin{align}
&\mathbb{P}\left(M^{\star}+\sum\limits_{k=1}^K\left((\xi_k-Q_k)_+-(\xi_k-Q_k-L^{\star}_k)_+\right)<S\right)=\frac{h+f+\lambda_o+\theta}{\lambda_u+\lambda_o+\theta} \label{FOC1}\\
&\mathbb{P}\left(M^{\star}+\sum\limits_{k=1}^K\left((\xi_k-Q_k)_+-(\xi_k-Q_k-L^{\star}_k)_+\right)<S \bigg| \xi_j>Q_j+L^{\star}_j\right)  %\notag\\&\qquad\qquad 
= \frac{ \frac{\theta}{\mathbb{P}(\xi_j>Q_j+L^*_j)} +\lambda_o-(h+r_j)}{\lambda_u+\lambda_o+\theta}, \nonumber\\
&\qquad\qquad  {\rm for}\qquad j=1,\dots, K. \label{FOC2}
\end{align}

\end{enumerate}

\end{myprop}

The optimality criterion \eqref{FOC1}-\eqref{FOC2} is simple - it depends on the probabilities of an execution shortfall  $A(X,\xi)<S$. Proposition 4 thus establishes that for $X$ to be optimal it is necessary and sufficient that $X$ equates these execution shortfall probabilities to  values computed with model parameters.

When the number of exchanges $K$ is large, shorfall probabilities in \eqref{FOC1}-\eqref{FOC2} are difficult to compute in closed-form. However, the case $K=2$ is relatively tractable and will be analyzed as an illustration. The assumption of independence between $\xi_1,\xi_2$ is made only in this example and is not required for the rest of our results. In Section \ref{sec.sa} we present results for correlated order flows.

\noindent {\bf Corollary}
{\it
In addition to assumptions of Proposition 4, assume that $K=2$ and $\xi_1,\xi_2$ are independent. Also, assume that $F_{1,2}(Q_{1,2})<1-\dfrac{h+r_{2,1}}{\lambda_o}$ and $(h+r_{1,2})\mathbb{P}\left(\xi_{1,2}>Q_{1,2}+S\right) > \theta$. Then there is an optimal order allocation $X^\star=(M^{\star}, L^{\star}_1, L^{\star}_2)\in int\{\mathcal{C}\}$ and it solves the following equations
\small
\begin{subequations}
\begin{align}
& L^{\star}_1  = Q_2 + S - M^{\star} - F^{-1}_2\left(\frac{-\theta/\bar F_1(Q_1+L^{\star}_1)+\lambda_u+\theta+h+r_1}{\lambda_u+\lambda_o+\theta}\right) \label{sol.2exchg.1}\\
& L^{\star}_2  = Q_1 + S -  M^{\star} - F^{-1}_1\left(\frac{-\theta/\bar F_2(Q_2+L^{\star}_2)+\lambda_u+\theta+h+r_2}{\lambda_u+\lambda_o+\theta}\right) \label{sol.2exchg.2}\\
& \bar F_1(  Q_1+L^{\star}_1)\bar F_2(Q_2+S-  M^{\star}-L^{\star}_1)
 +\int\limits_{Q_1+ S-M^{\star}-L^{\star}_2}^{Q_1+L^{\star}_1}{\bar F_2(Q_1+Q_2+S-M^{\star}-x_1)dF_1(x_1)}= \frac{\lambda_u-(h+f)}{\lambda_u+\lambda_o+\theta}\label{sol.2exchg.m},
\end{align}
\end{subequations}
\normalsize
\noindent where $F_1(\cdot), F_2(\cdot)$ are the cdf functions of $\xi_1,\xi_2$ respectively and $\bar F_i=1-F_i$.
}

\noindent In  \eqref{sol.2exchg.1}--\eqref{sol.2exchg.2} the optimal limit order quantities $L^{\star}_1,L^{\star}_2$ are linear functions of  $M^{\star}$. When \eqref{sol.2exchg.1}-\eqref{sol.2exchg.2} are substituted into \eqref{sol.2exchg.m} we obtain a non-linear equation for $M^{\star}$, which  can be solved for $M^*$.

\section{An optimal order routing algorithm}\label{sec.sa}
\subsection{A stochastic algorithm based on order flow sampling}

Practical applications require a fast and flexible method for optimizing order placement across multiple trading venues. As the number of venues increases, it becomes progressively more difficult to evaluate the objective function $V(X)$ - a $K$-dimensional integral - and to obtain analytical solutions for the order placement problem. In this section we propose a stochastic approximation method for  efficiently computing the optimal allocation even in high dimensions. The idea is to sample  order outflows $\xi_k$ and approximate the gradient of $V(X)$ along a random optimization path. Applying this approach for our problem formulation yields an intuitive iterative algorithm that updates trader's order allocation in response to past order execution outcomes.

Our numerical solution is based on the robust stochastic approximation algorithm of \cite{Nemirovski2009}. Denote by $g(X,\xi)=\nabla v(X,\xi)$ the gradient of $v(X,\xi)$  with respect to $X$. 
 The idea is to use a random samples  $\xi^n\in\mathbb{R}^{K}$ from $\xi$ to approximate, iteratively, $V(X)$ and its gradient:
\vskip 6pt
\begin{algorithmic}[1]
\STATE Choose an initial $X_0\in\mathbb{R}^{K+1}$ and fix a step size $\gamma$;
\FOR{$n=1,\dots,N$}
\STATE   $X_{n}=X_{n-1}-\gamma g(X_{n-1},\xi^n)$
\ENDFOR
\STATE \textbf{end}
\STATE An estimator $X^*$ is given by: $\hat X_N^\star=\frac{1}{N}\sum_{n=1}^NX_n$
\end{algorithmic}
\vskip 6pt

\noindent Here random variables $\xi^n$ are assumed to be an ergodic sequence sampled from the distribution $F$, which may or may not be known. Under  mild assumptions, satisfied by our objective function, the estimator $\hat X_N^\star$ converges to an optimal solution $X^\star$ and has a performance bound \newline $V(\hat X^\star)-V(X^\star)\leq\frac{DM}{\sqrt{N}}$, where $D=\underset{X,X'\in\mathcal{C}}{\max}\|X-X'\|_2$, $M=\sqrt{\underset{X\in\mathcal{C}}{\max}~\mathbb{E}\left[\|g(X,\xi)\|^2_2\right]}$.

\noindent The optimal step size is $\gamma^\star=\frac{D}{\sqrt{N}M}$ and we use a step size

$$\gamma=K^{1/2}S\left(N(h+f+\theta+\lambda_u+\lambda_o)^2 +N\sum\limits_{k=1}^K(h+r_k+\theta+\lambda_u+\lambda_o)^2\right)^{-1/2}$$

\noindent that scales appropriately with problem parameters. For  more details on stochastic approximation methods we refer to \cite{Kushner2003} and \cite{Nemirovski2009}.

In general, this method requires a sample of size $N$ of the random variable $\xi$. Since
$g(X_n,\xi)$ depends on random variables $\xi$ only through indicator functions:
\small
$$
g(X_n,\xi)=
\left(
\begin{matrix}
& h + f +\theta -(\lambda_u+\theta)\indicator{A(X_n,\xi)<S} + \lambda_o\indicator{A(X_n,\xi)>S}\\
\\
&\theta + \indicator{\xi_1>Q_1+L_{1,n}}\bigl(-(h+r_1) -(\lambda_u+\theta)\indicator{A(X_n,\xi)<S} + \lambda_o\indicator{A(X_n,\xi)>S}\bigr) \\
&\\
&\dots\\
&\\
&\theta + \indicator{\xi_K>Q_K+L_{K,n}}\bigl(-(h+r_K) -(\lambda_u+\theta)\indicator{A(X_n,\xi)<S} + \lambda_o\indicator{A(X_n,\xi)>S}\bigr) \\
\end{matrix}
\right)
$$
\normalsize
\vskip 6pt
\noindent a practical approach is to store the values of these indicator functions from past order executions at each exchange and use them to compute the sample values  $\xi^n$.

These indicators have simple interpretations: $\indicator{A(X_n,\xi)<S}=1$ if on the $n-$th order submission the trader fell behind the target quantity, $\indicator{A(X_n,\xi)>S}=1$ if he was ahead of the target, and $\indicator{\xi_k>Q_k+L_{k,n}}=1$ if a limit order on exchange $k$ was fully executed. This leads to a non-parametric online implementation of the algorithm, which updates order sizes in response to previous order execution outcomes, and can be interpreted as a sequential learning procedure. For example, the first row of  $g(X_n,\xi)$ describes updates of the market order quantity $M$:
\begin{itemize}
\item on each iteration $M$ is decreased by $\gamma(h + f+\theta)$ to reduce trading costs and market impact
\item if a trader fell behind his target quantity, $M$ will increase by $\gamma(\lambda_u+\theta)$ to reduce the shortfall on the next execution
\item since overtrading is also penalized, $M$ is decreased by $\gamma\lambda_o$ whenever a target quantity is exceeded
\end{itemize}
\noindent Limit order sizes are updated similarly. If a limit order is not filled, its quantity will be reduced by $\gamma\theta$ to reduce market impact, otherwise the update depends on a fill outcome.

The algorithm is fast - it updates an allocation vector sequentially and each update involves at most $5(K+1)$ arithmetic operations. On a retail laptop computer, an optimal order allocation across 12 exchanges can be computed in less than 200 milliseconds using an approximation with $N=1000$ iterations. The sequential nature of the algorithm allows to easily adapt it for low-latency trading applications: each iteration takes a fraction of a millisecond. First, optimal allocations are computed off-line and stored in a table. The table can be indexed by parameters such as $Q_k, S, T$. Then, an allocation is retrieved from the table in real-time, routing decisions are made and orders are sent. Finally, order fills are collected, the allocation is updated by one iteration as described by $g$, and stored in the table for future use. In Section \ref{ssec.tick} we present an illustration of this approach based on historical tick data.

\subsection{Numerical convergence}

\begin{figure}[h!]
\noindent
\begin{center}
{
\includegraphics[width=140mm]{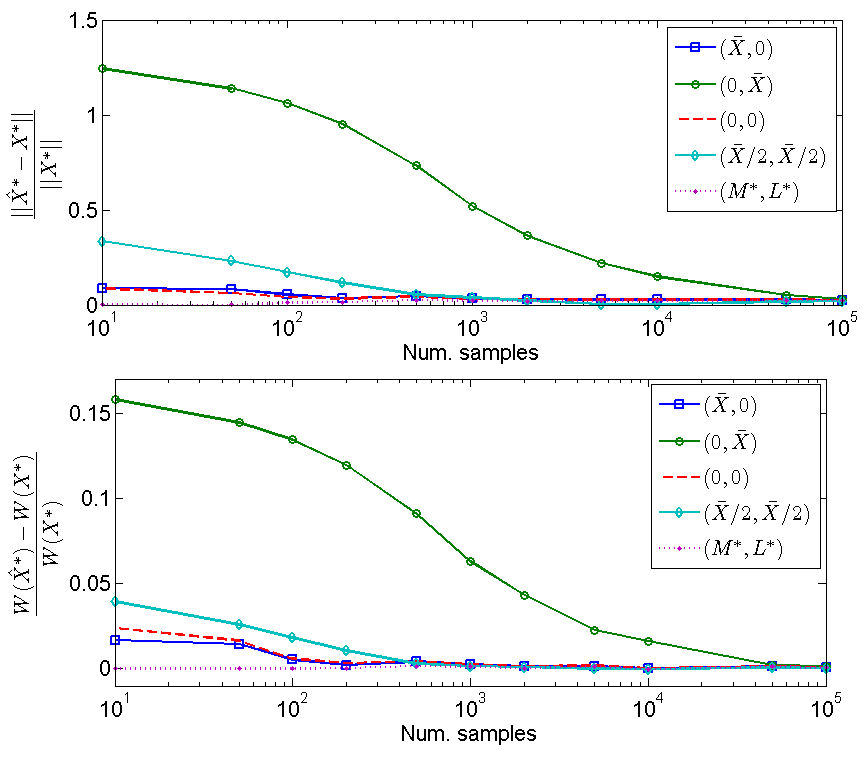}
}
\end{center}
\caption{Convergence of numerical solutions and objective values to an optimal point from different initial points.}
\label{fig.convergence_x}
\end{figure}

The robust stochastic approximation algorithm has a theoretical convergence rate of $\sqrt{N}$ as shown in \cite{Nemirovski2009}, i.e. $V(X^\star_N)-V(X^\star)\leq\frac{C}{\sqrt{N}}$, where $V(X^\star_N)$ is the objective function at an approximate solution $X^\star_N$ computed with $N$ iterations. When applied to our problem, this algorithm typically converges to an optimal point after 1,000-10,000 iterations for a given set of parameter values. To illustrate this, we applied it to an example with $K=1$ venue, and compared its results with a closed-form solution \eqref{sol.1exchg}. Using the same parameter values as on Figure \ref{fig.Sl_onex} with $\xi\sim Pois(\mu T)$, $\mu=2200$ shares per minute and $T=1$ minute the optimal solution is $(M^{\star},L^{\star})=(730,270)$ shares.  Numerical solutions $\hat X^\star$ were then computed for five starting points $X_0$ with a progressively larger number of iterations $N$. For each choice of $X_0$ and $N$ we also estimated an average cost per share $W(\hat X^\star)$ using an additional $L=1000$ samples of $\xi$ generated after $\hat X^\star$ is estimated. Figure \ref{fig.convergence_x} shows that numerical solutions converge to $X^\star$ regardless of the initial point $X_0$ and moreover starting from an optimal point with $X_0=X^\star$ the iterates remained close to this point. Convergence is quite fast - for most choices of the initial point the algorithm is within 2\% of the optimal objective value after only 50 iterations.

\newpage
\section{Numerical results}\label{sec.num}
\subsection{Comparative statics}\label{ssec.compstat}

To gain insights into the structure of order placement and routing decisions we turn to comparative statics analysis. Varying one model parameter at a time, we recompute an optimal allocation using the numerical algorithm from Section \ref{sec.sa} and plot optimal order sizes for different parameter values.  We also compare them to analytical solutions for $K=1$ obtained with same parameters.

Several solution features come into view in this analysis. First, we notice that in most cases it is optimal to oversize the total quantity of limit orders placed on multiple exchanges. Since limit order executions at individual exchanges are random, placing larger orders on multiple venues reduces the overall execution risk. This tendency to ``overbook" with multiple limit orders gradually vanishes as market impact increases. Similarly, it becomes less useful to oversize orders as order flows become more correlated across exchanges, decreasing the diversification advantage of having multiple orders. The oversizing becomes more prominent as the number of trading venues grows leading to more opportunities to diversify limit order fills. Larger target trade sizes also lead to more order oversizing. This suggests that large traders need to seek liquidity opportunities to fill a large trade even at a cost of creating some market impact with their limit orders.

Second, we find that limit orders have a bounded capacity for executing large trades. The quantity that is likely to be filled with a limit order on exchange $k$ is given by a queue size $Q_k$ and a order outflow distribution, more precisely by its tail $\mathbb{P}(\xi_k>x), x>Q_k$. Filling a relatively small quantity $S$ can usually be achieved by just placing several limit orders, but to fill larger quantities a trader needs to rely on market orders. A number of available venues plays a role in this tradeoff - the more venues are available, the more likely a trader can fill some of his limit orders, which makes market orders less necessary.

Third, we show that the target quantity $S$ itself determines which parameters drive its optimal allocation. For smaller $S$, our solutions quickly shift to limit orders on venues that provide the largest effective rebate. For larger $S$ we see that under- and overfill penalties play a more significant role and solutions become insensitive to rebates. Queue sizes and order outflow distributions appear to be important in all cases.

The baseline set of parameters in our analysis is representative of a typical stock in the U.S. equity market. We set $f=0.003, h=0.02, \theta=0.0005, \lambda_u=\lambda_o=0.05$ and $K=2$. Exchange parameters are set to $Q_k=2000, r_k=0.002$ for all venues. Order outflows follow a simple single-factor model, capturing the fact that order flows in a fragmented market are usually positively correlated. Specifically we put $\xi_k~=~\alpha\xi_0~+~(1-\alpha)\epsilon_k$, where $\xi_0,\epsilon_k$ are i.i.d. Poisson random variables with a common mean parameter $\mu T$, $\mu = 2200$ shares/minute, $T=1$ minute and $\alpha=0.6$. To compute each numerical solution we used $N=1000$ samples. This was enough for convergence, regardless of the number of exchanges $K$ as the algorithm remains largely unchanged with different $K$. With the exception of a comparison across values of $\mu$, in all other solution comparisons we used the same samples of $\xi_k$ across a range of parameter values. For analytical solutions we assumed a simple Poisson distribution for order flows with the same mean $\mu T$. Comparing solutions for $K=1$ and $K=2$ we find that order allocations for multiple venues are in many cases similar to their single-venue counterparts.

\begin{figure}[h!]
\noindent
\begin{center}
{
\includegraphics[width=170mm]{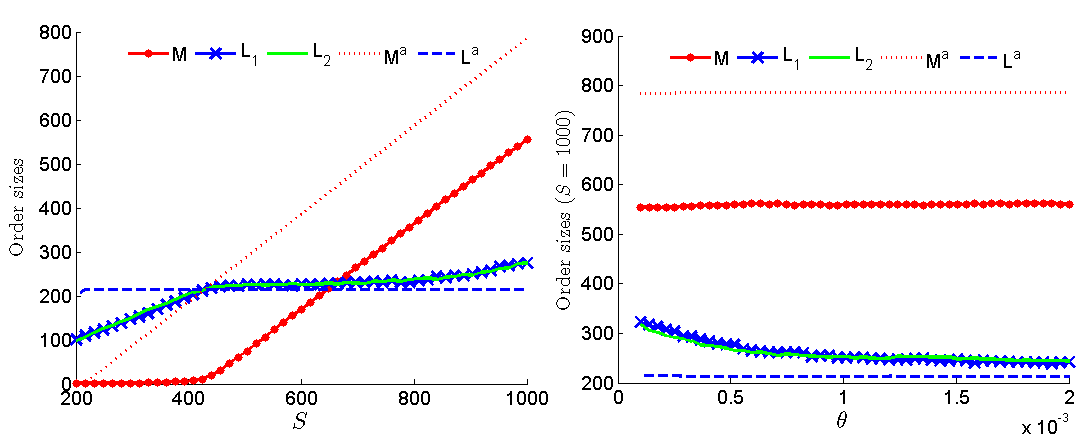}
}
\end{center}
\caption{Optimal order sizes with two exchanges $(M,L_1,L_2)$ and one exchange $(M^a,L^a)$, left panel: across a range of target quantities $S$, right panel: across a range of market impact coefficients $\theta$}
\label{fig.sens_S_theta}
\end{figure}

First, we compare solutions across a range of target quantities $S$. The left panel on Figure \ref{fig.sens_S_theta} shows that market orders can be avoided to fill small quantities ($S\leq 400$ in this example), but after a certain point an optimal market order size increases linearly with $S$ whereas limit order sizes stay relatively constant. Total order quantities $M^\star + L^\star_1 +  L^\star_2$ are larger than $S$, especially for bigger $S$. The right panel shows how this excess size decreases due to larger market impact that makes it more costly to oversize orders. For a single exchange there is no benefit in oversizing orders and according to Proposition \ref{prop.1exchg} $M^a + L^a = S$.

\begin{figure}[h!]
\noindent
\begin{center}
{
\includegraphics[width=170mm]{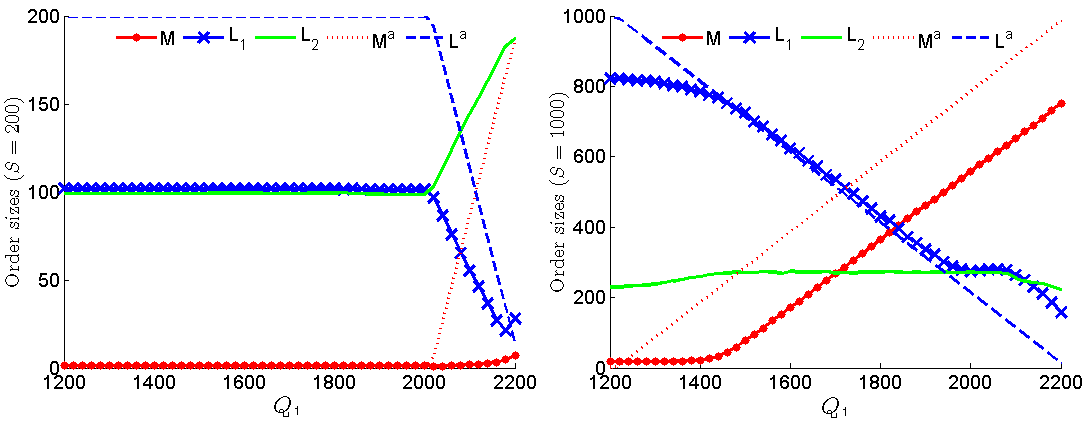}
}
\end{center}
\caption{Optimal order sizes with two exchanges $(M,L_1,L_2)$ and one exchange $(M^a,L^a)$ across a range of queue sizes $Q_1$, left panel: small trade $S=200$ shares, right panel: larger trade $S=1000$ shares}
\label{fig.sens_Q1}
\end{figure}

The second set of plots compares order allocations for different queue sizes. We distinguish between two scenarios - a target quantity $S=200$ which is small relative to our queue size and order flow distribution parameters and a large target quantity $S=1000$. Increasing the queue size at an exchange decreases the chance of filling a limit order there and creates a neccessity to find a substitute for that order. The left panel on Figure \ref{fig.sens_Q1} shows that when $S$ is small the trader can keep up with his execution using just limit orders even when the queue size at exchange 1 increases. The only difference it makes is that he places all orders on exchange 2 instead of splitting them between two venues. The right panel shows a different scenario - when $S$ is large the trader cannot substitute his limit orders as easily. All of the capacity for limit orders on exchange 2 is already used by his 200-share limit order, and he is forced to substitute his limit orders on exchange 1 with more expensive market orders.

The third set of plots compares solutions for different order outflow rates $\mu$. We again distinguish between a small and a large trade quantity $S$ but find that both cases are similarly affected by order flow rates. When $\mu$ is larger, i.e. there are more market orders, execution prospects improve for limit orders on all exchanges at the same time. More limit orders are used and the share of market orders in the optimal solution decreases.

\begin{figure}[h!]
\noindent
\begin{center}
{
\includegraphics[width=170mm]{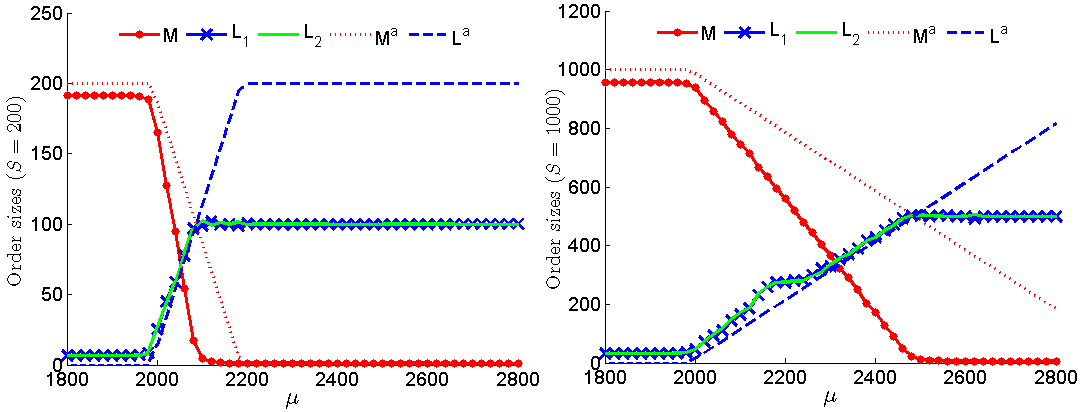}
}
\end{center}
\caption{Optimal order sizes with two exchanges $(M,L_1,L_2)$ and one exchange $(M^a,L^a)$ across a range of order flow rates $\mu$, left panel: small trade $S=200$ shares, right panel: larger trade $S=1000$ shares}\label{fig.sens_mu}
\end{figure}

\begin{figure}[h!]
\noindent
\begin{center}
{
\includegraphics[width=170mm]{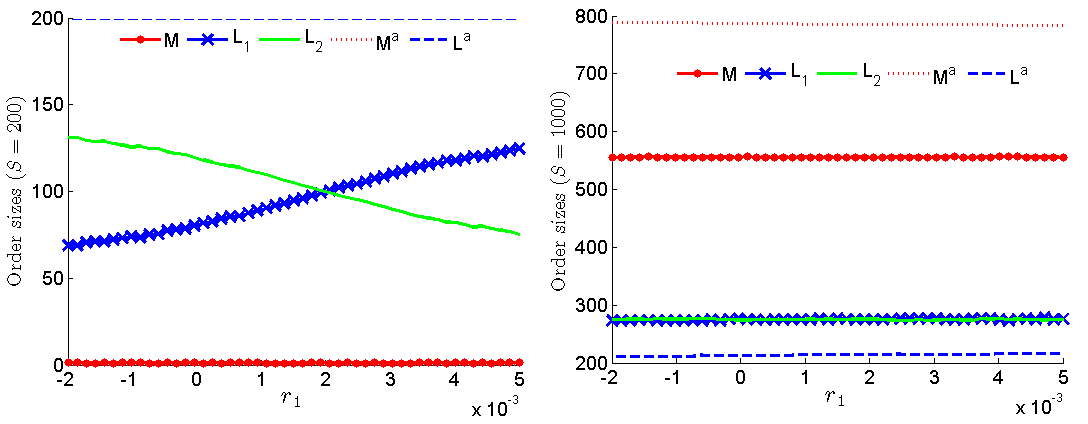}
}
\end{center}
\caption{Optimal order sizes with two exchanges $(M,L_1,L_2)$ and one exchange $(M^a,L^a)$ across a range of rebates $r_1$, left panel: small trade $S=200$ shares, right panel: larger trade $S=1000$ shares}
\label{fig.sens_r1}
\end{figure}

We also compare optimal solution across a range of pricing parameters - effective rebates $r$, execution shortfall penalties $\lambda_u,\lambda_o$, fee $f$ and bid-ask spread $h$. Figure \ref{fig.sens_r1} shows that traders with small execution quantities favor exchanges with larger rebates (net of adverse selection). Small quantities can be filled with limit orders at any exchange. Traders that have this flexibility prefer to capture a larger rebate. In contrast, when the execution quantity is large traders need to use limit orders on all venues. In this case the optimal solution is insensitive to rebates because limit orders at any exchange are less costly than market orders and the trader needs to use limit orders at all exchanges to the maximum extent to fill a large quantity.

\begin{figure}[h!]
\noindent
\begin{center}
{
\includegraphics[width=170mm]{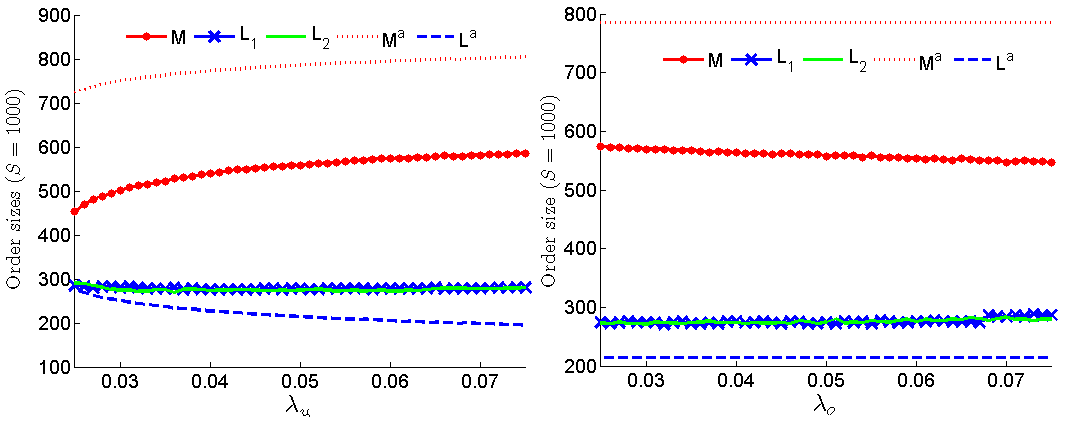}
}
\end{center}
\caption{Optimal order sizes with two exchanges $(M,L_1,L_2)$ and one exchange $(M^a,L^a)$, left panel: across a range of $\lambda_u$, right panel: across a range of $\lambda_o$}
\label{fig.sens_lambdas}
\end{figure}

\begin{figure}[h!]
\noindent
\begin{center}
{
\includegraphics[width=170mm]{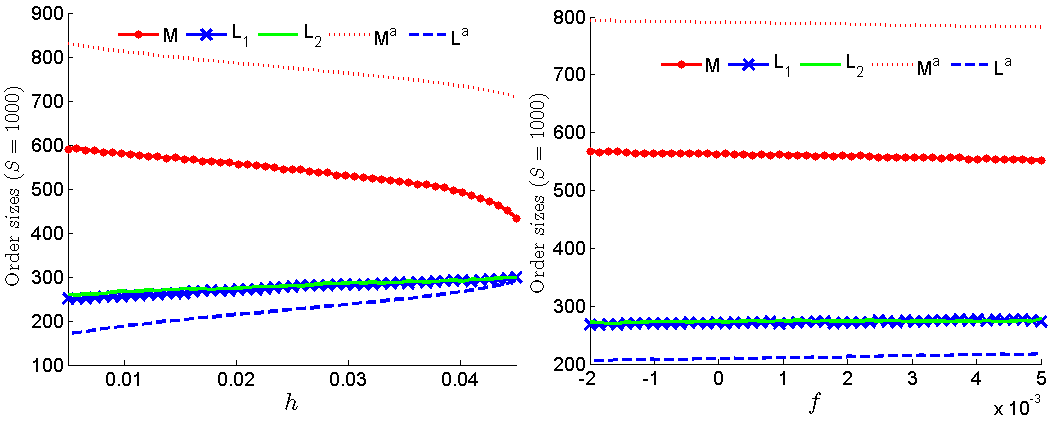}
}
\end{center}
\caption{Optimal order sizes with two exchanges $(M,L_1,L_2)$ and one exchange $(M^a,L^a)$, left panel: across a range of spreads $h$, right panel: across a range of fees $f$}
\label{fig.sens_h_f}
\end{figure}

Optimal solutions for small trade quantities are practically unaffected by changes in $\lambda_u,\lambda_o, h$ and $f$. For small trades market orders can be avoided and there is little execution risk which renders these parameters unimportant. For large target quantities these parameter play a larger role. Figures \ref{fig.sens_lambdas} and \ref{fig.sens_h_f} show that more market orders are used when underfilling an order is expensive, and less market orders are used when these orders themselves become more costly due to large spread or fee. The dependence of market order sizes on $\lambda_o$ is also intuitive - as overfills become more expensive the trader will use more limit orders instead of market orders because they have a smaller risk of overfilling and compensate for this risk with a lower cost.

Lastly, we compare optimal solutions for a different number of exchanges and estimate trading costs for our optimal solution. The optimal cost is compared to that of a few simpler strategies: a pure market order allocation $X^M = (S,0,\dots,0)$, a single limit order allocation $X^L=(0,S,0,\dots,0)$ and an equal split allocation $X^E = \left(\dfrac{S}{K+1},\dfrac{S}{K+1},\dots,\dfrac{S}{K+1}\right)$.

Optimal order allocations presented in Table 1 suggest that the best strategy is to use limit orders to the maximum possible extent, which depends on a trade quantity. As the number of exchanges increases, the optimal allocation shifts from market orders to limit orders for all trade sizes, and for smaller sizes it shifts entirely. Market orders do not need to be used at all if there are enough exchanges for the trader to place limit orders on. The tendency to place more limit orders than needed is also pronounced in these examples, especially for larger trades which require the trader to use all opportunities to fill his limit orders.

Our analysis of execution costs for different order placement strategies shows a clear benefit in optimizing the tradeoff between market orders and limit orders in a fragmented market. When multiple exchanges are available the difference between a naive strategy and an optimal solution is 2-4 cents per share, which is economically significant. More importantly our results show that simple static strategies cannot accomodate different execution sizes. For example, the pure market order strategy $X^M$ performs relatively well for large $S$ but it is too expensive for small trades which can be executed more efficiently with limit orders. Similarly, a trader can achieve relatively low costs by equally splitting his target quantity into multiple orders if the total size is small, but this rule performs even worse than $X^M$ for a large sizes. The optimal proportion of market and limit orders as well as their allocation across exchanges varies significantly based on trade characteristics and market dynamics.

\newpage
\begin{table}[!h]
\begin{center}
\small
\begin{tabular}{|r|r r r r r r r|r r r r|}
\hline
\multirow{2}{*}{$K$} & \multicolumn{7}{|c|}{Order allocation in \% of $S$} & \multicolumn{4}{|c|}{Average cost, in cents per share}\\
\cline{2-12}
 & $\hat M^\star$ & $L^\star_1$ & $L^\star_2$ & $L^\star_3$ & $L^\star_4$ &  $\hat L^\star_5$ & Total & $W(X_M)$ & $W(X_L)$ & $W(X_E)$ & $W(X^\star)$\\
\hline
\multicolumn{12}{|c|}{$S=500$}\\
\hline
1 & 82\% & 18\% &  &  &  &  & 100\% & 2.35 & 2.19 & 0.83 & 1.54\\

2 & 15\% & 44\% & 44\% &  &  &  & 103\% & 2.35 & 2.22 & -0.57 & -0.85\\

3 & 1\% & 34\% & 34\% & 34\% &  &  & 102\% & 2.35 & 2.21 & -1.02 & -1.99\\

4 & 0\% & 26\% & 25\% & 26\% & 26\% &  & 102\% & 2.35 & 2.20 & -1.25 & -2.06\\

5 & 0\% & 22\% & 21\% & 20\% & 20\% & 20\% & 103\% & 2.35 & 2.22 & -1.40 & -2.05\\
\hline
\multicolumn{12}{|c|}{$S=1000$}\\
\hline
1 & 94\% & 6\% &  &  &  &  & 100\% & 2.35 & 3.65 & 2.27 & 2.07\\

2 & 56\% & 27\% & 27\% &  &  &  & 111\% & 2.35 & 3.64 & 1.28 & 0.77\\

3 & 35\% & 23\% & 23\% & 23\% &  &  & 104\% & 2.35 & 3.65 & 0.09 & -0.07\\

4 & 15\% & 23\% & 23\% & 22\% & 23\% &  & 106\% & 2.35 & 3.64 & -0.88 & -0.90\\

5 & 1\% & 21\% & 21\% & 21\% & 21\% & 21\% & 106\% & 2.35 & 3.65 & -1.29 & -1.64\\
\hline
\multicolumn{12}{|c|}{$S=5000$}\\
\hline
1 & 97\% & 3\% &  &  &  &  & 100\% & 2.35 & 4.81 & 3.44 & 2.22\\

2 & 88\% & 8\% & 8\% &  &  &  & 104\% & 2.35 & 4.81 & 3.60 & 2.10\\

3 & 83\% & 9\% & 9\% & 9\% &  &  & 110\% & 2.35 & 4.81 & 3.55 & 1.95\\

4 & 79\% & 11\% & 11\% & 11\% & 11\% &  & 124\% & 2.35 & 4.81 & 3.39 & 1.79\\

5 & 75\% & 11\% & 11\% & 11\% & 11\% & 11\% & 129\% & 2.35 & 4.82 & 3.22 & 1.62\\
\hline
\end{tabular}
\normalsize
\vskip 4pt
\caption{Left panel: optimal order allocations, right panel: costs for the optimal allocation and simpler benchmarks}
\end{center}
\end{table}

\begin{table}[!h]
\begin{center}
\small
\begin{tabular}{|r|r r r r|r r|r r|}
\hline
\multirow{2}{*}{$K$} & \multicolumn{4}{|c|}{Costs, cents per share} & \multicolumn{2}{|c|}{Average shortfall, shares} & \multicolumn{2}{|c|}{Shortfall probabilities}\\
\cline{2-9}
 & Fees & Impact & Penalties & Total & Underfill & Overfill & Underfill & Overfill\\
\hline
\multicolumn{9}{|c|}{$S=500$}\\
\hline
1 & 1.49 & 0.05 & 0.00 & 1.54 & 0 & 0 & 100\% & 0\%\\
2 & -1.35 & 0.06 & 0.44 & -0.85 & 40 & 4 & 71\% & 29\%\\
3 & -2.18 & 0.05 & 0.14 & -1.99 & 6 & 9 & 20\% & 80\%\\
4 & -2.23 & 0.05 & 0.11 & -2.06 & 0 & 11 & 3\% & 97\%\\
5 & -2.24 & 0.05 & 0.14 & -2.05 & 0 & 14 & 1\% & 100\%\\
\hline
\multicolumn{9}{|c|}{$S=1000$}\\
\hline
1 & 2.02 & 0.05 & 0.00 & 2.07 & 0 & 0 & 100\% & 0\%\\
2 & 0.41 & 0.06 & 0.30 & 0.77 & 51 & 9 & 74\% & 26\%\\
3 & -0.48 & 0.06 & 0.36 & -0.07 & 64 & 7 & 73\% & 27\%\\
4 & -1.38 & 0.06 & 0.42 & -0.90 & 71 & 12 & 67\% & 34\%\\
5 & -2.07 & 0.06 & 0.38 & -1.64 & 55 & 21 & 51\% & 49\%\\
\hline
\multicolumn{9}{|c|}{$S=5000$}\\
\hline
1 & 2.17 & 0.05 & 0.00 & 2.22 & 2 & 0 & 100\% & 0\%\\
2 & 1.85 & 0.05 & 0.19 & 2.10 & 192 & 0 & 100\% & 0\%\\
3 & 1.66 & 0.06 & 0.23 & 1.95 & 230 & 0 & 99\% & 1\%\\
4 & 1.46 & 0.06 & 0.26 & 1.79 & 256 & 1 & 98\% & 2\%\\
5 & 1.30 & 0.07 & 0.26 & 1.62 & 257 & 2 & 96\% & 4\%\\
\hline
\end{tabular}
\normalsize
\vskip 4pt
\caption{Order placement costs and fill statistics for the optimal allocation}
\end{center}
\end{table}
\newpage

\subsection{Application to tick data}\label{ssec.tick}

To further illustrate our method, we applied it to historical tick data using a specific trade example. We considered an execution of a moderate-sized order to buy $S=2000$ shares of Microsoft Corporation (MSFT) stock with an execution deadline $T=1$ minute, to be traded on two exchanges - NASDAQ and BATS Z. This liquid stock is traded on multiple exchanges, but for simplicity we considered only these two. We assumed in this simulation that rebates are $r_{NSDQ}=0.2$ and $r_{BATS}=0.25$ cents per share which is close to their historical averages. The fee was assigned to $f=0.29$ cents per share and the half-spread was set to $h=0.50$ cents per share which is typical for this stock. To perform our numerical optimization we used trade and quote (TAQ) data from January to March 2012, and then analyzed its performance on a dataset from April 2012.

\begin{figure}[h!]
\noindent
\begin{center}
{
\includegraphics[width=140mm]{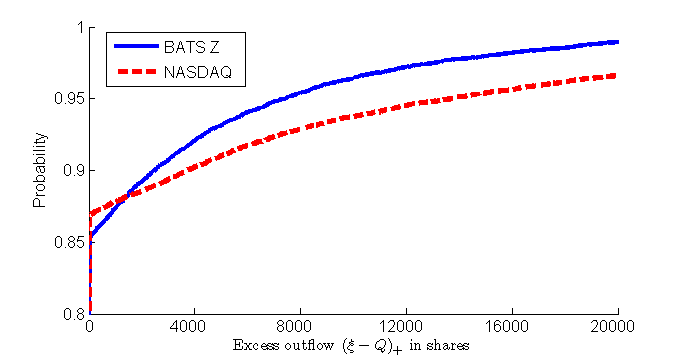}
}
\end{center}
\caption{Empirical cumulative distribution functions of market order flows in excess of initial queue size}
\label{fig.data_cdf}
\end{figure}

Average queue sizes on NASDAQ and BATS Z in the calibration dataset were equal to 12,392 and 8,179 shares respectively, and average 1-minute market sell order volumes for each exchange were equal to 848 and 922 shares. These averages however do not reflect the tails of a volume distribution. Figure \ref{fig.data_cdf} shows that in about 15\% of observations the volume of market sell orders during an interval $[0,T]$ was larger than the queue size at time 0, often by thousands of shares. In these cases limit orders placed at the back of a queue at time 0 will be filled and our theoretical results show that the likelihood of these tail events drives optimal limit order placement decisions.

Our numerical optimization does not require to estimate or specify order flow distributions. Instead, as described in Section \ref{sec.sa} it requires data on limit order fills, which were simulated in this example using tick data. To simulate a limit order fill we assumed that a buy limit order $L$ placed behind a queue of $Q$ shares at the best bid is filled if the volume of sell market orders during its placement horizon ($T=1$ minute in our example) is larger than $Q$. The execution is complete if the volume of sell market orders exceeds $Q$ by more than $L$, otherwise it is a partial execution. One-minute sell market order volumes were estimated as a sum of trade sizes at a given exchange whose trade prices were equal to the prevailing bid price. In this simulation we made several simplifications that are related to TAQ data limitations. First, our limit order fill estimates are conservative as they do not include possible order cancelations from the front of a bid queue. TAQ data does not have information on order cancelations, but this information is available in more detailed ``level-2" datasets. Second, buy limit orders can be filled when ask price ticks down, but estimating these fills accurately requires knowledge of order queue position which is unavailable in TAQ. To simplify our simulation we restricted it to one-minute samples where the best quotes did not change up or down.

To estimate sensitivity to queue sizes $Q_{BATS}$ or $Q_{NSDQ}$ we ranked observations by queue sizes $Q_{BATS},Q_{NSDQ}$ and grouped them into three equal sized bins labeled ``high", ``medium" and ``low". Similarly we ranked, grouped and labeled observations by total trading volume in the previous minute $VOL_{BATS},VOL_{NSDQ}$. Each label was then used as a state variable leading to $3^4 = 81$ different combinations of state variables. Each combination of states defined a subsample of historical data, which we used to simulate limit order fills. Based on simulated limit order fills we computed a numerical solution for each subsample following the algorithm from Section \ref{sec.sa}.

The result is a lookup table where one can find an approximate solution for each of 81 state variable combinations. Order allocations in this table take into account the magnitude of $Q_{BATS},Q_{NSDQ}$ and $VOL_{BATS},VOL_{NSDQ}$ relative to their historical values. We recomputed order allocations using this method for a number of parameter values and Table 1 presents a subset of these solutions for different levels of  $VOL_{BATS}$ and $Q_{BATS},Q_{NSDQ},VOL_{NSDQ}$ all set to their corresponding ``low" states.

\begin{table}[!h]
\begin{center}
\begin{tabular}{|l|l|r r r r|}
\hline
\multirow{2}{*}{Shortfall penalties $\lambda_u, \lambda_o$} & \multirow{2}{*}{$VOL_{BATS}$} & \multicolumn{4}{|c|}{Order allocation in \% of $S$}\\
\cline{3-6}
 &  & $M^\star$ & $L^\star_{BATS}$ & $L^\star_{NSDQ}$ & Total\\
\hline
\multirow{3}{*}{1.0 cent per share} & low & 99\% & 1\% & 1\% & 100\%\\
 & medium & 98\% & 2\% & 1\% & 101\%\\
 & high & 4\% & 93\% & 8\% & 105\%\\
\hline
\multirow{3}{*}{0.9 cents per share} & low & 13\% & 82\% & 82\% & 176\%\\
 & medium & 3\% & 95\% & 59\% & 157\%\\
 & high & 3\% & 96\% & 4\% & 103\%\\
\hline
\multirow{3}{*}{0.5 cents per share} & low & 0\% & 99\% & 99\% & 199\%\\
 & medium & 0\% & 99\% & 99\% & 199\%\\
 & high & 0\% & 100\% & 21\% & 121\%\\
\hline
\end{tabular}
\caption{Optimal order allocations conditioned on a state variable $VOL_{BATS}$}
\end{center}
\end{table}

\begin{table}[!h]
\begin{center}
\begin{tabular}{|r|r r|r r|}
\hline
\multirow{2}{*}{Shortfall penalties $\lambda_u, \lambda_o$} & \multicolumn{2}{|c|}{Shares filled} & \multicolumn{2}{|c|}{Cost}\\
\cline{2-5}
 & $X^\star$ & $X_E$ & $X^\star$ & $X_E$\\
\hline
1.0 cent per share & 1469 & 867 & 0.85 & 0.84\\
0.9 cents per share & 564 & 867 & 0.74 & 0.78\\
0.5 cents per share & 526 & 867 & 0.38 & 0.55\\
\hline
\end{tabular}
\caption{Average execution quantity and cost (in cents per share) based on a tick data simulation}
\end{center}
\end{table}

This example demonstrates how an optimal order allocation shifts from  limit orders to market orders as a trader's aversion to execution risk (represented by $\lambda_u,\lambda_o$) increases. We also notice that an order allocation depends on $VOL_{BATS}$ - since trading volume is positively autocorrelated in time, a ``high" reading of volume in the previous minute predicts a higher volume in the next minute, making a limit order execution more likely on the BATS Z exchange. Therefore the numerical solution places more limit orders there. Interestingly, the optimal allocation puts limit orders on both NASDAQ and BATS Z when $VOL_{BATS}$ is ``medium", but when it is ``high" almost no orders are placed on NASDAQ because a BATS execution is already quite  likely.

The average execution costs for our numerical solution was computed based on a separate data sample and compared to the costs of an equal split strategy $X_E = \left(\frac{S}{3},\frac{S}{3},\frac{S}{3}\right)$. Despite a substantial simplification of the original optimization problem the approximate solution based on states outperforms a static benchmark. Similarly to the previous section we find that a static approach is too inflexible - for example when execution risk aversion parameters are relatively low $\lambda_u=\lambda_o=0.5$ cents per share, a better execution can be achieved by using fewer market orders. Since more limit orders are used in our strategy, its average fill size 526 is lower than to 867 shares obtained with $X_E$. However, our solution leads to a lower cost because execution risk parameters are relatively low and overall we can achieve a lower cost by using more limit orders.

The computation presented here can be viewed only as a first step towards implementing an order routing system. There is a variety of practical issues that were not addressed, such as fine-tuning risk aversion parameters $\lambda_u, \lambda_o$ or defining meaningful state variables given intraday changes in distributions of queue sizes and order flows. Still, we are hopeful that results and intuition developed here will be useful for designing order routing systems.

\newpage
\section{Conclusion}\label{sec.conc}

We have formulated the {\it optimal order placement problem} for a market participant able to submit market orders and limit orders across multiple exchanges, and studied its solution properties using historical data and numerical simulations. In the case when only one exchange is available we have derived an optimal split between limit and  market orders and showed that an optimal order allocation depends on trader's aversion to execution risk. For the general case of multiple exchanges, we provide a characterization of the optimal order placement strategy in terms of execution shortfall probabilities. To solve the problem in practical applications we propose a fast and straightforward numerical algorithm that re-samples past order fill data to optimize future order executions. Using this algorithm, we have studied the sensitivities of an optimal order allocation to problem inputs and showed that a simultaneous placement of limit orders on multiple trading venues according to our method can lead to a substantial reduction of transaction costs.

% Appendix here
% Options are (1) APPENDIX (with or without general title) or
%             (2) APPENDICES (if it has more than one unrelated sections)
% Outcomment the appropriate case if necessary
%

\begin{APPENDIX}{A: Alternative problem formulation}

\noindent We may also consider an alternative approach to order placement optimization, which turns out to be related to our original formulation by duality. Consider the following problem:
\begin{myproblem}[Alternative formulation: cost minimization under execution constraints]
\begin{align}
&\min_{X\in\mathbb{R}^{K+1}_+}\mathbb{E}\left[(h+f)M -\sum_{k=1}^K(h+r_k)((\xi_k-Q_k)_+-(\xi_k-Q_k-L_k)_+)\right. \notag\\ &\qquad\qquad\left.+\theta\left(M+\sum_{k=1}^KL_k+\left(S - A(X,\xi)\right)_+\right)\right]\label{optim.set.constr}\\
&\text{subject to:}\quad\mathbb{E}\left[\left(S - A(X,\xi)\right)_+\right]\leq\mu_u \label{optim.constr.1}\\
&\hskip 60pt \mathbb{E}\left[\left(A(X,\xi) - S\right)_+\right]\leq\mu_o
\label{optim.constr.2}
\end{align}
\end{myproblem}

\noindent In this alternative formulation a trader can specify his tolerance to execution risks using constraints on expected order shortfalls and overflows. The goal is to minimize an expectation of order execution costs under expected shortfall constraints. The Problem 2 does not appear to be tractable, but it has a convex objective and convex inequality constraints, and we can easily find its (Lagrangian) dual problem:

\begin{myproblem}
\begin{equation}\label{optim.set.dual}
\max_{\lambda_u\geq 0,\lambda_o\geq 0}\left\{V^\star(\lambda_u,\lambda_o)-\lambda_u\mu_u-\lambda_o\mu_o\right\}
\end{equation}
where $V^\star(\lambda_u,\lambda_o) = \underset{X\in\mathbb{R}^{K+1}_+}{\min}\mathbb{E}[v(X,\xi)]$ is the optimal objective value from Problem 1 given parameter values $\lambda_u,\lambda_o$.
\end{myproblem}

\noindent We see that the Problem 3 is related to our original order placement problem - solving the Problem 3 (and therefore, the Problem 2) amounts to re-solving the Problem 1 for different values of $\lambda_u,\lambda_o$. This discussion also leads to a new interpretation of parameters $\lambda_u,\lambda_o$ in the Problem 1 as shadow prices for expected shortfall and overflow constraints in the related Problem 2. Hereafter we focus on the (more tractable) Problem 1, but note that the optimal point for the Problem 2 can also be found by solving its dual problem.

\end{APPENDIX}

\begin{APPENDIX}{B: Proofs}

\noindent{\bf Proof of Proposition \ref{prop.lubound}}
First, for any allocation $\tilde X$ that has $\tilde M>S$, we automatically have $A(\tilde X)>S$ and we can show that the (random) costs of $\tilde X$ are larger than those of $X^M=(S,0,\dots,0)\in\mathcal{C}$:

\begin{multline}
v({\tilde X},\xi) - v(X^M,\xi)=
(h+f)(\tilde M-S)-\sum_{k=1}^K(h+r_k)((\xi_k-Q_k)_+-(\xi_k-Q_k-\tilde L_k)_+)+\\ \lambda_o\left(\tilde M-S+\sum_{k=1}^K\left((\xi_k-Q_k)_+-(\xi_k-Q_k-\tilde L_k)_+\right)\right) +\theta( M-S+\sum_{k=1}^K\tilde L_k) > \\
(\lambda_o+h+f)(\tilde M-S)+\sum_{k=1}^K(\lambda_o-h-r_k)((\xi_k-Q_k)_+-(\xi_k-Q_k-L_k)_+)>0, \notag
\end{multline}
\noindent which holds for all random $\xi$. Therefore, $V(\tilde X)>V(X^M)$.
Similarly, for any allocation $\tilde X$  with $\tilde L_k>S - \tilde M$ define a new allocation $\tilde X'$ by setting $\tilde M'=\tilde M$, $\tilde L'_j=\tilde L_j, \forall j\neq k$ and $\tilde L'_k=S - \tilde M$. Then $v(\tilde X,\xi) - v(\tilde X',\xi)=\theta(\tilde L_k - \tilde L'_k) > 0$ on the event $B~=~\left\{\omega|\xi_k(\omega)<Q_k+S - M\right\}$.

On a complementary event $B^c$,
$$v(\tilde X,\xi)-v(\tilde X',\xi)=-(h+r_k)((\xi_k-Q_k-S+\tilde M)_+-(\xi_k-Q_k-\tilde L_k)_+)$$
$$+\lambda_o((\xi_k-Q_k-S+\tilde M)_+-(\xi_k-Q_k-\tilde L_k)_+)+\theta(\tilde L_k - \tilde L'_k).$$

Therefore
 \begin{multline}
V(\tilde X) - V(\tilde X') =
\mathbb{E}\left[v(\tilde X,\xi)- v(\tilde X',\xi)|B\right]\mathbb{P}(B)+
\mathbb{E}\left[v(\tilde X,\xi)-v(\tilde X',\xi)|B^c\right]\mathbb{P}(B^c) = \\ \theta(\tilde L_k - \tilde L'_k)+\mathbb{E}\left[(\lambda_o-(h+r_k))((\xi_k-Q_k-S+\tilde M)_+-(\xi_k-Q_k-\tilde L_k)_+)|B^c\right]\mathbb{P}(B^c)> 0 \notag
\end{multline}

\noindent If $\tilde X'\notin\mathcal{C}$, we can continue truncating limit order sizes $\tilde L'_j>S - \tilde M'$ to $S - \tilde M'$ following the same argument. Each time the truncation decreases the objective function and finally we obtain a $\tilde X''\in\mathcal{C}$, such that $V(\tilde X'') < V(\tilde X)$.

Next, if $\tilde X$ is such that $\tilde M-\sum_{k=1}^K{\tilde L_k}<S$ consider the difference $s=S-\tilde M-\sum_{k=1}^K{\tilde L_k}$ and define a new allocation $\tilde X'$ as $\tilde M'=\tilde{M}+s, \tilde L'_k=\tilde{L}_k, k=1,\dots,K$. Then $v(\tilde X',\xi)-v(\tilde X,\xi) = (h + f - \lambda_u)s < 0$. Although the total quantity of orders in allocation $\tilde X$ is smaller than in $\tilde X'$, it has the same market impact. The former always falls short of the target $S$ and its remainder $s$ must be filled with a market order at time $T$, leading to the same impact. Since $v(\tilde X',\xi)-v(\tilde X,\xi) <0$, we also have $V(\tilde X') < V(\tilde X)$.  $\blacksquare$
\newline

\noindent \textbf{Proof of Proposition \ref{prop.objfun}:}
First, note that $(\xi_k-Q_k)_+-(\xi_k-Q_k-L_k)_+$ are concave functions of $L_k$. Therefore, $A(X,\xi)$ is concave as a sum of concave functions. Similarly, the cost term in $v(X,\xi)$ is a sum of convex functions, as long as $r_k\geq -h, k=1,\dots,K$ and is itself a convex function. Second, since $S - A(X,\xi)$ is a convex functon of $X$, and the function $h(x)=\lambda_u(x)_+-\lambda_o(-x)_+$ is convex in $x$ for positive $\lambda_u,\lambda_o$, so the penalty term $h\left(S - A(X,\xi))\right)$ is also convex.

Using the assumption $\lambda_o > h+\underset{k}{\max}\{r_k\}$ it is clear that the function $v(X,\xi)$ is bounded from below by $-(h+\underset{k}{\max}\{r_k\})S$ and therefore $V(X)$ is also bounded from below.

Since $V(X)$ is convex, it is also continous and since it is bounded from below it reaches a local minimum $V_{min}$ on the compact set $\mathcal{C}$ at some point $X^\star\in\mathcal{C}$. By convexity, $V_{min}$ is a global minimum of $V(X)$ on $\mathcal{C}$. Moreover, Proposition \ref{prop.lubound} guarantees that $V_{min}<V(\tilde X)$ for any $\tilde X\notin\mathcal{C}$, so $V_{min}$ is also a global minimum of $V(X)$ on  $\mathbb{R}^{K+1}_+$. $\blacksquare$
\newline

\noindent{\bf Proof of Proposition \ref{prop.1exchg}:}

 By Proposition \ref{prop.lubound} there exists an optimal split $ (M^\star,L^\star)\in\mathcal{C}$ between limit and market orders. Moreover for $K=1$ the set $\mathcal{C}$ reduces to a line $M^\star+L^\star=S$ so it is sufficient to find $M^\star$. The restriction $L=S-M$ implies that $\{A(X,\xi)>S\}=\varnothing$, $\{A(X,\xi)<S, \xi>Q+L\}=\varnothing$, and we can rewrite the objective function as
\begin{multline}\label{obj.onex}
V(M)=\mathbb{E}\Bigl[(h+f)M-(h+r)((\xi-Q)_+-(\xi-Q-S+M)_+) + \theta S~~+\\
(\lambda_u+\theta)\left(S - M-((\xi-Q)_+-(\xi-Q-S+M)_+)))\right)_+\Bigr].
\end{multline}
\noindent Although $\theta S$ is a constant, we note that different solutions $(M,L)$ have different total market impact due to a terminal market order at time $T$. For $M\in(0,S)$ the expression under the expectation in (\ref{obj.onex}) is bounded for all $\xi$ and Lipschitz with respect to $M$, so we can compute $V'(M)=\frac{dV(M)}{dM}$ as:
\begin{align}
V'(M)&=\mathbb{E}\Bigl[h+f+(h+r)\indicator{\xi>Q+S-M}-(\lambda_u+\theta)\indicator{\xi<Q+S-M}\Bigr] \notag \\
&=2h+f+r-(h+r+\lambda_u+\theta)F(Q+S-M)
\end{align}
\noindent Note that if $\lambda_u\leq\dfrac{2h+f+r}{F(Q+S)}-(h+r+\theta)$, then  $V'(M)\geq 0$ for $M\in(0,S)$ and therefore $V$ is non-decreasing at these points. Checking that $$V(S)-V(0)\geq(h+f-\lambda_u-\theta)S+(\lambda_u+\theta+h+r)S(1-F(Q+S))\geq 0$$ we conclude that $M^\star = 0$. Similarly, if $\lambda_u\geq\dfrac{2h+f+r}{F(Q)}-(h+r+\theta)$, then $V'(M)\leq 0$ for all $M\in(0,S)$ and $V(M)$ is non-increasing at these points. Checking that $$V(S)-V(0)\leq(h+f-\lambda_u-\theta)S+(\lambda_u+\theta+h+r)S(1-F(Q))\leq 0$$ we conclude that $M^\star = S$. Finally, if $\lambda_u$ is between these two values, $\exists\epsilon>0$, such that $V'(\epsilon)<0, V'(S-\epsilon)>0$ and by continuity of $V'$ there is a point where $V'(M^\star)=0$. This $M^\star$ is optimal by convexity of $V(M)$ and (\ref{sol.1exchg}) solves equations $V'(M^\star)=0, L^\star=S-M^\star$. $\blacksquare$
\newline

\noindent{\bf Proof of Proposition \ref{prop.foc}:  }

  Proposition \ref{prop.objfun} implies the existence of an optimal order allocation $X^\star\in\mathcal{C}$. First, we define $X^M=(S,0,\dots,0)$ and prove that $X^\star\neq X^M$ by contradiction. If $X^M$ were optimal in the problem (\ref{optim.set}) it would also be optimal in the same problem with a constraint $L_k=0,k\neq j$, for any one $j$. In other words, the solution $(S, 0)$ would be optimal for any one-exchange problem, defined by using only exchange $j$. But by our assumption, there exists a $J$ such that $\lambda_u<\dfrac{2h+f+r_J}{F_J(Q_J)}-(h+r_J)$ and Proposition \ref{prop.1exchg} implies that $(S, 0)$ is not optimal for the $J$-th single-exchange subproblem, leading to a contradiction.

The function $v(X,\xi)$ is bounded for  $X\in\mathcal{C}$ and for all $\xi$,  differentiable with respect to $M$ and $L_k, k=1,\dots,K$ for $X\in\mathcal{C}\backslash\left\{X^M\right\}$ for almost all $\xi$. Therefore  $V(X)$ is differentiable for $X\in\mathcal{C}\backslash\left\{X^M\right\}$ and we can compute all of its partial derivatives by interchanging the order of differentiation and integration. The KKT conditions for problem (\ref{optim.set}) and $X\in\mathcal{C}\backslash\left\{X^M\right\}$ are
\begin{align}
&h + f +\theta -(\lambda_u+\theta)\mathbb{P}(A(X^\star,\xi)<S) + \lambda_o\mathbb{P}(A(X^\star,\xi)>S)-\mu_0=0 \label{KKT1}\\
&-(h+r_k)\mathbb{P}(\xi_k>Q_k+L^{\star}_k)+\theta -(\lambda_u+\theta)\mathbb{P}(A(X^\star,\xi)<S, \xi_k>Q_k+L^{\star}_k) + \notag\\
& \qquad \lambda_o\mathbb{P}(A(X^\star,\xi)>S, \xi_k>Q_k+L^{\star}_k)-\mu_k=0,~~k=1,\dots, K \label{KKT2}\\
& M\geq 0,~~L_k\geq 0,~~\mu_0\geq 0,~~\mu_k\geq 0,~~ \mu_0M=0,~~\mu_kL_k=0,~~k=1,\dots, K \label{KKT3}
\end{align}
Since the objective function $V(\cdot)$ is convex, conditions \eqref{KKT1}--\eqref{KKT3} are both necessary and sufficient for optimality. The  first result of this proposition follows if we consider any $\tilde X$ with $\tilde M=0$ and use the assumption $\lambda_u\geq\frac{2h+f+\underset{k}{\max}\{r_k\}}{p_0}-(h+\underset{k}{\max}\{r_k\})$:

$$V(\tilde X)\geq-(h+\underset{k}{\max}\{r_k\})S\mathbb{P}\left(\overline{\underset{k}{\bigcap}\{\xi_k\leq Q_k\}}\right)+\lambda_u S\mathbb{P}\left(\underset{k}{\bigcap}\{\xi_k\leq Q_k\}\right)+\theta S\geq(h+f+\theta)S=V(X^M).$$

\noindent We already argued that $\exists X^\star$ with $V(X^\star)<V(X^M)$, so $X^\star\neq\tilde X$ and therefore $M^\star>0$.

Next, we can rearrange terms in a $j$-th equality (\ref{KKT2}) as follows:

\begin{equation}\label{KKT2supp}
\theta+\mathbb{P}(\xi_j>Q_j+L^{\star}_j)\left[\lambda_o-(h+r_j) -(\lambda_u+\lambda_o+\theta)\mathbb{P}(A(X^\star,\xi)<S | \xi_j>Q_j+L^{\star}_j)\right]-\mu_j=0.
\end{equation}

\noindent Setting $L^{\star}_j=0$ and using the assumption  $(h+r_j-\lambda_o)\mathbb{P}\left(\xi_j>Q_j\right) > \theta$ we see that equation \eqref{KKT2supp} cannot hold because the left-hand side is negative, leading to a contradiction.

We showed that $M^{\star}>0, L^{\star}_j>0$ for all $j=1,\dots,K$ and therefore, $\mu_0=\mu_1=\dots=\mu_K=0$ by complementary slackness. Then the KKT conditions \eqref{KKT1}--\eqref{KKT3} reduce to \eqref{FOC1}--\eqref{FOC2}.
$\blacksquare$

\noindent{\bf Corollary}
{\it
In addition to assumptions of Proposition 4, assume that $K=2$ and $\xi_1,\xi_2$ are independent. Also, assume that $F_{1,2}(Q_{1,2})<1-\dfrac{h+r_{2,1}}{\lambda_o}$ and $(h+r_{1,2})\mathbb{P}\left(\xi_{1,2}>Q_{1,2}+S\right) > \theta$. Then there is an optimal order allocation $X^\star=(M^{\star}, L^{\star}_1, L^{\star}_2)\in int\{\mathcal{C}\}$ and it solves equations (\ref{sol.2exchg.1}-\ref{sol.2exchg.m}).
}\newline

\noindent {\bf Proof:}
First we show that solutions on the boundary of $\mathcal{C}$ are sub-optimal: $M^\star = 0$, $M^\star = S$ and $L^\star_{1,2} = 0$ are ruled out in Proposition 4. Solutions with $M^\star+\sum\limits_{k=1}^KL^\star_k = S$ are ruled out by checking \eqref{KKT2} and noting that since $A(X^\star,\xi)>S=\varnothing$ and $(h+r_j)\mathbb{P}\left(\xi_j>Q_j+S\right) > \theta$ by assumption, the left-hand side of that equation is always negative, leading to a contradiction.

Finally, solutions with $L_{1,2}=S-M$ are also ruled out by checking \eqref{KKT2}. For example, if $L_1=S-M>0$ it follows that $\mu_1=0$ and we can rewrite \eqref{KKT2} as:

$$-(h+r_1)\mathbb{P}(\xi_1>Q_1+L_1)+\theta +\lambda_o\mathbb{P}(\xi_2>Q_2)\mathbb{P}(\xi_1>Q_1+L_1) = 0$$

Dividing by $\mathbb{P}(\xi_1>Q_1+L_1)$ and applying the assumption $F_1(Q_1)<1-\dfrac{h+r_2}{\lambda_o}$ we note that the left-hand side of the equation is always positive leading to a contradiction.

Having showed that an optimal solution belongs to the interior of $\mathcal{C}$, we can simplify the description of an event $A(X,\xi)>S$. For any $X\in int\{\mathcal{C}\}$, $A(X,\xi)>S$ if and only if all the following three inequalities are satisfied:

\begin{subequations}
\begin{align}
& \xi_1 > Q_1 + S - M - L_2  \label{overfill.2exchg.1}\\
& \xi_2 > Q_2 + S - M - L_1  \label{overfill.2exchg.2}\\
& \xi_1 + \xi_2 > Q_1 + Q_2 + S - M \label{overfill.2exchg.3}
\end{align}
\end{subequations}

\noindent These inequalities give a simple characterization of the event $\{A(X,\xi)>S\}$ which is directly verified by considering subsets of $(\xi_1,\xi_2)$ forming a complete partition of $\mathbb{R}^2_+$.

{\bf Case 1:} $\xi_1>Q_1+L_1, \xi_2>Q_2+L_2$. Since $L_1+L_2+M>S$, we  have $A(X,\xi)=L_1+L_2+M>S$ and at the same time all of the inequalities (\ref{overfill.2exchg.1}-\ref{overfill.2exchg.3}) are satisfied, so they are trivially equivalent in this case.

{\bf Case 2:} $\xi_1>Q_1+L_1, Q_2\leq\xi_2\leq Q_2+L_2$. Because of the condition $\xi_1>Q_1+L_1$, (\ref{overfill.2exchg.1}) is satisfied. We have in this case that $A(X,\xi)=L_1+\xi_2-Q_2+M$ and thus $A(X,\xi)>S$ if and only if (\ref{overfill.2exchg.2}) is satisfied. Finally, $\xi_1>Q_1+L_1$ together with (\ref{overfill.2exchg.2}) imply (\ref{overfill.2exchg.3}), so $A(X,\xi)>S$ and (\ref{overfill.2exchg.1}-\ref{overfill.2exchg.3}) are equivalent in this case.

{\bf Case 3:} $\xi_2>Q_2+L_2, Q_1\leq\xi_1\leq Q_1+L_1$. Similarly to Case 2 we can show that inequalities (\ref{overfill.2exchg.1}-\ref{overfill.2exchg.3}) are satisfied if and only if $A(X,\xi)>S$.

{\bf Case 4:} $Q_1+S-M-L_2<\xi_1\leq Q_1+L_1, Q_2+S-M-L_1<\xi_2\leq Q_2+L_2$. This set is non-empty because $0<S - M- L_1 <L_2$ and similarly for $L_1, L_2$ reversed. Inequalities \eqref{overfill.2exchg.1}--\eqref{overfill.2exchg.2} hold trivially, only (\ref{overfill.2exchg.3}) needs to be checked. We can write $A(X,\xi)=\xi_1-Q_1+\xi_2-Q_2+M>S$ if and only if (\ref{overfill.2exchg.3}) holds, so $A(X,\xi)>S$ is equivalent to (\ref{overfill.2exchg.1}-\ref{overfill.2exchg.3}).

{\bf Case 5:} Outside of Cases 1-4, either (\ref{overfill.2exchg.1}) or (\ref{overfill.2exchg.2}) is not satisfied. If $\xi_1\leq Q_1+S-M-L_2, \xi_2\leq Q_2+L_2$, then $A(X,\xi)\leq S-M-L_2 + L_2 +M = S$. The case $\xi_2\leq Q_2+S-M-L_1, \xi_1\leq Q_1+L_1$ is completely symmetric, and it shows that neither $A(X,\xi)> S$ nor (\ref{overfill.2exchg.1}-\ref{overfill.2exchg.3}) hold in this case.

\noindent Next, we use inequalities (\ref{overfill.2exchg.1}-\ref{overfill.2exchg.3}) to  characterize the set $\{A(X,\xi)>S\}$ in the first-order conditions \eqref{FOC1}--\eqref{FOC2}. We observe that in the two-exchange case
\begin{align*}
& \{A(X,\xi)>S,\xi_1>Q_1+L_1\}=\{\xi_1>Q_1+L_1,\xi_2>Q_2+S-M-L_1\}\\
& \{A(X,\xi)>S,\xi_2>Q_2+L_2\}=\{\xi_2>Q_2+L_2,\xi_1>Q_1+S-M-L_2\},
\end{align*}
\noindent and then use the independence of $\xi_1$ and $\xi_2$ to compute
\begin{align*}
& \mathbb{P}(A(X,\xi)>S|\xi_1>Q_1+L_1)=\bar F_2(Q_2+S-M-L_1) \\
&\mathbb{P}(A(X,\xi)>S|\xi_2>Q_2+L_2)=\bar F_1(Q_1+S-M-L_2)
\end{align*}
\noindent Together with \eqref{FOC2}, this leads to a pair of equations for limit orders sizes:
\begin{align*}
 \bar F_2(Q_2+S-M-L_1)=\frac{-\theta/\bar F_1(Q_1+L_1)+\lambda_u+\theta+h+r_1}{\lambda_u+\lambda_o+\theta} \\
\bar F_1(Q_1+S-M-L_2)=\frac{-\theta/\bar F_2(Q_2+L_2)+\lambda_u+\theta+h+r_2}{\lambda_u+\lambda_o+\theta}
\end{align*}
\noindent whose solution  is given by $L^{\star}_1, L^{\star}_2$ from (\ref{sol.2exchg.1},\ref{sol.2exchg.2}). To obtain the equation (\ref{sol.2exchg.m}), we rewrite the first equation in (\ref{FOC1},\ref{FOC2}) using the inequalities (\ref{overfill.2exchg.1}-\ref{overfill.2exchg.3}). Then ${P}(A(X,\xi)>S)$ may be computed as
the integral of the product measure $F_1 \otimes F_2 $ over the region defined by
$$ U(Q,S,M,L_1,L_2)= \{  (x_1,x_2)\in \mathbb{R}^2,\quad x_1>Q_1+ S - M-L_2,\quad x_2>Q_2+ S - M-L_1,\quad x_1+x_2>Q_1+Q_2+S-M \}.$$
This integral is given by
\begin{align*}
&{P}(A(X,\xi)>S)=F_1 \otimes F_2  \left( U(Q,S,M,L_1,L_2)\right)\\
&=\bar F_1( Q_1+L_1)\bar F_2(Q_2+S-M-L_1) +\int\limits_{Q_1+ S-M-L_2}^{Q_1+L_1}{\bar F_2(Q_1+Q_2+S-M-x_1)dF_1(x_1)}=\frac{\lambda_u-(h+f)}{\lambda_u+\lambda_o+\theta}
\end{align*}

$\blacksquare$

\end{APPENDIX}
%\newpage
%
%   or
%
% \begin{APPENDICES}
% \section{<Title of Section A>}
% \section{<Title of Section B>}
% etc
% \end{APPENDICES}

% Acknowledgments here
%\ACKNOWLEDGMENT{The authors gratefully acknowledge the existence of
%the Journal of Irreproducible Results and the support of the Society
%for the Preservation of Inane Research.}

\newpage

% References here (outcomment the appropriate case)

% CASE 1: BiBTeX used to constantly update the references
%   (while the paper is being written).
\bibliographystyle{ormsv080} % outcomment this and next line in Case 1
\bibliography{refs_routing} % if more than one, comma separated

% CASE 2: BiBTeX used to generate mypaper.bbl (to be further fine tuned)
%\input{mypaper.bbl} % outcomment this line in Case 2

%If you don't use BiBTex, you can manually itemize references as shown below.

%%%%%%%%%%%%%%%%%
\end{document}